\documentstyle [preprint,aps]{revtex}
\begin{document}
\draft

\title
{Quantum-Classical Correspondence via Liouville Dynamics: I. Integrable
Systems and the Chaotic Spectral Decomposition}
\author{Joshua Wilkie and Paul Brumer}
\address{Chemical Physics Theory Group, Department of Chemistry,
   University of Toronto,
   Toronto, Ontario, Canada M5S 1A1 }

\maketitle				      

\begin{abstract}
A general program to show quantum-classical correspondence
for bound conservative integrable and chaotic systems is described. The
method is applied to integrable systems and the nature of the approach to
the classical limit, the cancellation of essential singularities, is
demonstrated. The application to chaotic systems requires an understanding of
classical Liouville eigenfunctions and 
a Liouville spectral decomposition, developed herein. General approaches to
the construction of these Liouville eigenfunctions and classical spectral
projectors in quantum and classical mechanics are discussed 
and are employed to construct Liouville eigenfunctions for classically
chaotic systems. Correspondence for systems whose classical analogs are
chaotic is discussed, based on this decomposition, in the
following paper\cite{chacor}.
\end{abstract}

\section{Introduction} \label{intro}

The correspondence principle requires the laws of quantum dynamics to
reproduce the predictions of classical mechanics when the energy and
mass of a dynamical system are very large with respect to Planck's
constant\cite{cp}. In other words the correspondence principle asserts
that the fundamental laws of physics which are valid on the atomic scale are
also responsible for the observed
classical dynamics of macroscopic systems. It is therefore surprising, 
considering
the important conceptual role of the correspondence principle that there is
little direct evidence to support its general validity \cite{ford2,ford,berry4}. In fact it has been
controversially\cite{berry4,bw1} argued that
the absence of chaos in quantum dynamics plus its presence in the
dynamics of classical systems prevents any possibility of
correspondence\cite{ford}. The main goal of this paper and its companion \cite{chacor} is to
provide a consistent theory of correspondence in both integrable and
chaotic systems.

To understand correspondence requires that we first adopt 
a framework in which quantum and classical dynamics can be
realistically compared.  The Liouville picture in classical
mechanics in
combination with a phase space representation of quantum
dynamics provides such a framework. Here both classical and quantum
mechanics are represented by objects defined on phase space and
by operators defined on these objects\cite{Koopman}.
 Our correspondence approach is then based on the desire to
prove that the quantum dynamics of phase space distributions obeys
classical mechanics in the classical limit. We do so by
focusing on the correspondence of what we argue to be the essential 
elements in dynamics, the eigendistributions and
eigenvalues of the classical and quantum Liouville operators\cite{dirac,dirac2}
(the Poisson bracket with the classical Hamiltonian $H$, and the
commutator with the Hamiltonian operator $\hat{H}$, respectively).

We outline this correspondence program in Section \ref{LPCP} and describe the
progress made and developments required in the application to regular
and chaotic
systems. The approach emphasizes the significance of 
the eigenfunctions of the classical Liouville operator and
of the classical spectral projectors. These eigenfunctions are obtained in
Section \ref{CLEE} for both regular and chaotic systems. 
The correspondence
program is then implemented for integrable systems in Section
\ref{corres} and
the nonsingular nature of the approach to the classical limit is
displayed. This approach to correspondence proves inappropriate for
chaotic systems. In this case a description of the dynamics in terms
of classical spectral projectors is required. These quantities are
introduced and discussed in Section \ref{CLSD}.   A companion paper\cite{chacor} shows classical-quantum
correspondence for chaotic systems.  

\section{Liouville Picture: The Correspondence Program}
\label{LPCP}

Arguments in favor of the Liouville picture as the conceptual
framework most suited for the study of correspondence have been
presented elsewhere\cite{jaffe2,jaffe1}. Here we briefly summarize.
First, the Liouville picture of classical dynamics, which deals with
the time evolution of phase space densities, is conceptually superior to
the trajectory viewpoint because it implicitly
recognizes the imperfect nature of classical preparation and measuring
devices. That is, one can not initially prepare the exact points in
 phase space which define a single
trajectory. Similarly, the
limitations of quantum mechanical preparation and measurement can be
 incorporated in the von
Neumann (quantum Liouville) equation for the density matrix, i.e.,
the von Neumann equation admits
solutions which are not states of maximal information. 
Thus, the Liouville equation plays a similar role in quantum and
classical dynamics. 
Second, our approach takes advantage of
the similarity between the dynamics of classical distributions in
phase space, as governed by the classical Liouville equation, and the
dynamics of quantum density matrices, governed by the von Neumann
equation in the Wigner-Weyl representation. Adopting this perspective,
both classical and quantum mechanics have similar structures and rely
upon similar quantities. 
Finally, formal classifications of classical systems as integrable, ergodic,
mixing, etc. are most properly done in terms of the eigenfunctions and
eigenvalues of the classical Liouville Operator\cite{arnold}, i.e., the Poisson
bracket with the Hamiltonian. Hence this provides the most fundamental
of frameworks.

The time evolution of the quantum density matrix $\hat{\rho}$ is given by
the solution to the von Neumann (or quantum Liouville) equation:
\begin{equation}
\frac{\partial \hat{\rho}}{\partial t}=-i\hbar^{-1}[\hat{H}, \hat{\rho}]=
-iL\hat{\rho},
\label{eq1}
\end{equation}
where $L=\hbar^{-1}[\hat{H},\cdot]$ is the quantum Liouville operator, and 
$[\cdot,\cdot]$ is the commutator.
Solutions to Eq. (\ref{eq1}) are of the form:
\begin{equation}
\hat{\rho}=\hat{\rho}_{\lambda,\alpha}e^{-i\lambda t}
\end{equation}
if $\hat{\rho}_{\lambda,\alpha}$ satisfies the eigenvalue problem
\begin{equation}
L\hat{\rho}_{\lambda,\alpha}= \lambda \hat{\rho}_{\lambda,\alpha}.
\label{quanteig}
\end{equation}
The label $\alpha$ is introduced to accommodate
degeneracies associated with the eigenvalue $\lambda$. In particular,
a complete characterization of the quantum Liouville eigenvalue
problem requires the specification of a complete set of commuting
superoperators
in the same way that solution of the Schrodinger eigenvalue problem
requires a complete set of commuting observables\cite{dirac}. This
complete
set of commuting superoperators can be constructed
in the following way. Suppose
\begin{equation}
\hat{H},\hat{K}_1,\dots,\hat{K}_{r'-1}
\label{CSCO}
\end{equation}
for $r'\leq s$ is a complete
set of observables for a system of $s$ degrees of freedom such that
$[\hat{H},\hat{K}_i]=0$ and $[\hat{K}_i,\hat{K}_j]=0$ for all $i,j\leq
r'-1$. It
follows that the complete set of $2r'$ commuting quantum superoperators is 
\begin{equation}
L,{\cal H},\frac{1}{\hbar}[\hat{K}_1,\cdot],\frac{1}{2}[\hat{K}_1,\cdot]_+,
\dots,\frac{1}{\hbar}[\hat{K}_{r'-1},\cdot],\frac{1}{2}[\hat{K}_{r'-1},\cdot]_+
\label{Lcal}
\end{equation}
where ${\cal
H}=\frac{1}{2}[\hat{H},\cdot]_+$ is the energy superoperator, and $[\cdot,\cdot]_+$
denotes the anticommutator, i.e.,
$[\hat{A},\hat{B}]_+=\hat{A}\hat{B}+\hat{B}\hat{A}$. A Liouville
eigenstate for the system is completely specified if it is an
eigenstate of all $2r'$ superoperators. Thus $\hat{\rho}_{\lambda,\alpha}$ 
in Eq. (\ref{quanteig}) is an eigenfunction of the collection of
superoperators in Eq. (\ref{Lcal}), with the label $\alpha$ 
denoting the eigenvalues of all superoperators other than $L$. 

Equations (\ref{eq1}) and (\ref{quanteig}) are of the following
form in the Wigner-Weyl representation\cite{ww} [${\bf x}=({\bf p},{\bf q})$ 
denotes the collection of momenta ${\bf p}$ and coordinates ${\bf q}$]
\begin{equation}
{\frac{\partial \rho^w({\bf x},t)}{\partial t}} =-iL({\bf x})\rho^w({\bf x},t) 
\end{equation}
and 
\begin{equation}
 L({\bf x})\rho^w_{\lambda,\alpha}({\bf x})=
 \lambda\rho^w_{\lambda,\alpha}({\bf x}),
 \label{extra}
 \end{equation}
where
\begin{equation}
L({\bf x})= \frac{2}{\hbar}i H({\bf x})\sin(\hbar\sigma/2)= 
\frac{2}{\hbar} iH({\bf x})\sin
\left[\frac{\hbar}{2}\left(
\frac{\stackrel{\leftarrow}{\partial}}{\partial q}
\frac{\stackrel{\rightarrow}{\partial}}{\partial p}- 
\frac{\stackrel{\leftarrow}{\partial}}{\partial p}
\frac{\stackrel{\rightarrow}{\partial}}{\partial q} \right) \right].
\label{qlopww}
\end{equation}
Here $\sigma$ implies the Poisson bracket, i.e.,
\begin{equation}
A({\bf x})\sigma B({\bf
x})=\{A,B\} =\frac{\partial A({\bf x})}{\partial {\bf q}}\cdot
\frac{\partial B({\bf x})}{\partial {\bf p}}-\frac{\partial A({\bf x})}
{\partial {\bf p}}\cdot
\frac{\partial B({\bf x})}{\partial {\bf q}}
\label{PoissonB}
\end{equation}
and the arrows over the derivatives indicate that the
derivative is taken of the function preceding or following the
derivative operator.

Given the solution to Eq. (\ref{extra}), the dynamics of any
distribution $\rho^w({\bf x},t)$, with initial condition 
$\rho^w({\bf x},t=0)=\rho({\bf x},0)$, can be written as:
\begin{equation}
\rho^{w}({\bf x},t)=\sum_{\lambda,\alpha} {c}_{\lambda,\alpha}
\rho^w_{\lambda,\alpha} ({\bf x}) e^{-i\lambda t}
\label{quantsum}
\end{equation}
where
\begin{equation}
{c}_{\lambda,\alpha}=\int d{\bf x}_0 \rho ({\bf x}_0,0)
\rho^{w*}_{\lambda,\alpha}({\bf x}_0).
\label{quantcoeff}
\end{equation}
That is,
\begin{equation}
\rho^{w} ({\bf x},t)=\int d{\bf x}_0 \rho ({\bf x}_0,0)
\left[\sum_{\lambda,\alpha}
\rho^{w}_{\lambda,\alpha}({\bf x}) 
\rho^{w*}_{\lambda,\alpha}({\bf x}_0)\right] e^{-i\lambda t}.
\label{qexp}
\end{equation}

Consider now the analogous approach in classical mechanics which
considers the dynamics of phase space distributions $\rho({\bf x},t)$.
In particular, $\rho({\bf x},t)$ satisfies the Liouville equation
\begin{equation}
\frac{\partial \rho({\bf x},t)}{\partial t}=-iL_c \rho({\bf x},t) 
\label{what}\end{equation}
where the classical Liouville operator $L_c$ is given by 
\begin{equation}
L_c=i\{H({\bf x}), \cdot \}.
\label{classL}
\end{equation}
Here $H(\bf x)$ is the Hamiltonian and $\{ \cdot,\cdot \}$ denotes the Poisson
bracket [Eq. (\ref{PoissonB})]. The time evolution of $ \rho({\bf x},t)$ is
given by
\begin{equation}
\rho({\bf x},t)=\rho_{\lambda^c,\alpha}({\bf x}) e^{-i\lambda^c t}
\label{sep}
\end{equation}
if $\rho_{\lambda^c,\alpha}({\bf x})$ satisfies the eigenvalue problem
\begin{equation}
L_c \rho_{\lambda^c,\alpha}({\bf x})=\lambda^c \rho_{\lambda^c,\alpha}({\bf x}).
\label{classeiq}
\end{equation}
Once again $\alpha$ contains the labeling associated with states which
are $\lambda^c$ degenerate.  In particular, 
a complete characterization of the classical
Liouville eigenvalue problem requires the specification of a complete
set of commuting operators on the classical Hilbert
space\cite{jaffe1,necnsuf}. Suppose the classical analog of our quantum
system has $r$ independent constants of the motion. Then we can
construct $r-1$ functions $K_j({\bf x})$ which are constants of the
motion, i.e., $\{H({\bf x}),K_j({\bf
x})\}=0$, and which are in involution so that
$\{K_i({\bf x}),K_j({\bf x})\}=0$ for all $i,j\leq r-1$.  The
complete set of commuting operators on the classical Hilbert space is then
\begin{equation}
L_c,H({\bf x}), i\{K_1({\bf x}),\cdot\},K_1({\bf x}), \dots,
i\{K_{r-1}({\bf x}),\cdot\},K_{r-1}({\bf x}),
\label{fifteen}
\end{equation}
where $L_c,i\{K_1({\bf x}),\cdot\},\dots,
i\{K_{r-1}({\bf x}),\cdot\}$ are
first order linear differential operators and $H({\bf x}),K_1({\bf
x}),\dots,K_{r-1}({\bf x})$ are multiplicative operators. A Liouville
 eigenstate for the classical problem is usually completely specified
if it is an eigenstate of all $2r$ classical operators\cite{necnsuf}. 

Given such solutions, the evolution of any
classical distribution $\rho({\bf x},t)$ can be written as
\begin{equation}
\rho({\bf x},t)=\sum_{\lambda^c,\alpha} c_{\lambda^c,\alpha}
~\rho_{\lambda^c,\alpha} ({\bf x}) e^{-i\lambda^c t}
\label{classsum}
\end{equation}
where
\begin{equation}
{c}_{\lambda^c,\alpha}=\int d{{\bf x}_0} ~\rho({\bf x_0},0) 
{\rho^*}_{\lambda^c,\alpha} (\bf x_0),
\label{classcoff}
\end{equation}
or
\begin{equation}
\rho({\bf x},t)= \int d{\bf x}_0 ~\rho({\bf x}_0,0)
\left[\sum_{\lambda^c,\alpha}\rho_{\lambda^c,\alpha} ({\bf x})
 {\rho^*}_{\lambda^c,\alpha} ({\bf x}_0) \right]
 e^{-i\lambda^c t}.
\label{cexp}
\end{equation}

Both Eqs. (\ref{qexp}) and (\ref{cexp}) express the time evolution of
the phase space density in terms of the initial distribution
$\rho({\bf x}_0,0)$, the eigenvalues $\lambda, \lambda^c$ and a term comprised
of products of Liouville eigenfunctions contained within brackets.
This rewriting of the right hand sides of the equations, and the term
in brackets in particular, will prove to play a central role in the
correspondence program we develop in these papers. Indeed the term in
brackets, in Eqs. (\ref{qexp}) and (\ref{cexp}), is related to the
Liouville spectral projection operators, which are heavily emphasized
in this paper.

A comparison of Eqs. (\ref{qexp}) and (\ref{cexp}) motivates our
correspondence approach. Specifically, it suggests that one can show
quantum-classical correspondence by showing that the $h\rightarrow 0$
limit of the quantum Liouville eigenfunctions and its eigenvalues are
the classical Liouville eigenfunctions and eigenvalues. This approach
has proven successful for integrable systems \cite{jaffe2}. However,
as shown below and in the following paper \cite{chacor}, it is an
incorrect approach for chaotic systems. In such cases the quantum
Liouville eigenstates have no classical limit. Rather, one must deal
with quantum and classical spectral projection operators, which have
singularity-free classical limits.  These are related to the objects
contained within brackets in Eq. (\ref{qexp}) and (\ref{cexp}).

The status of this correspondence program is readily summarized for
integrable systems ($r=r'=s$).
Berry\cite{berry1} demonstrated correspondence for the stationary
Liouville eigenfunctions $|{\bf n}\rangle\langle {\bf n}|$ where
$|{\bf n}\rangle $ are eigenstates of the Hamiltonian operator with
corresponding energies $E_{{\bf n}}$ and the $s$ integers ${\bf n}$
label the quantum states of the Schrodinger picture. Specifically,
he showed that
\begin{equation}
\rho_{{\bf n},{\bf n}}^w({\bf x})\equiv h^{-s/2}\int d{\bf
v}e^{i{\bf p}\cdot{\bf v}/\hbar}\langle {\bf q}-{\bf
v}/2| {\bf n}\rangle\langle {\bf n}|{\bf q}+{\bf v}/2\rangle\rightarrow
\hbar^{s/2}\rho_{{\bf I}_{\bf n},0}({\bf x})
\end{equation}
in the $h \rightarrow 0$ limit, where ${\bf I}_{\bf
n}=({\bf n}+\mbox{\boldmath $\beta$})\hbar$,
the constants $\mbox{\boldmath $\beta$}$ are the Maslov
indices\cite{jaffe2,delos}, $\rho_{{\bf n},{\bf n}}^w$ is the
Wigner transform of the stationary Liouville eigenstate $|{\bf
n}\rangle\langle {\bf n}|$, and the classical Liouville 
eigenfunctions $\rho_{{\bf I}',{\bf k}}({\bf x})$ are defined below 
[Eq. (\ref{cefs})].
Correspondence for
the nonstationary Liouville eigenfunctions $|{\bf
n}\rangle\langle {\bf m}|$ and eigenvalues $\lambda_{{\bf n},{\bf
m}}=(E_{{\bf n}}-E_{{\bf m}})/\hbar$ was considered by Jaff\'{e} and
Brumer\cite{jaffe2}. In particular, they examined the Wigner
transform $\rho_{{\bf n},{\bf m}}^w({\bf x})$ of $|{\bf n}\rangle
\langle{\bf m}|$ and showed that for ${\bf n}\neq{\bf m}$
\begin{equation}
\rho_{{\bf n},{\bf m}}^w({\bf x})\equiv h^{-s/2}\int d{\bf
v}e^{i{\bf p}\cdot{\bf v}/\hbar}\langle {\bf q}-{\bf
v}/2| {\bf n}\rangle\langle {\bf m}|{\bf q}+{\bf v}/2\rangle\rightarrow
\hbar^{s/2}\rho_{{\bf I}_{{\bf n},{\bf
m}},{\bf n}-{\bf m}}({\bf x}),
\end{equation}
and
\begin{equation}
\lambda_{{\bf n},{\bf m}}\rightarrow \lambda^c_{{\bf I}_{{\bf n},{\bf
m}},_{{\bf n}-{\bf m}}}
\end{equation}
as $h\rightarrow 0$,
where they surmised that ${\bf I}_{{\bf n},{\bf m}}=({\bf I}_{\bf
n}+{\bf I}_{\bf m})/2$. However, as we will later
show, this ${\bf I}_{{\bf n},{\bf m}}$ is an approximation to the
exact result which depends intimately on the eigenvalues of the complete set
of commuting observables [Eq. (\ref{CSCO})]. These classical Liouville
eigenvalues $\lambda^c_{{\bf I}',_{{\bf k}}}$ are defined below 
[Eq. (\ref{rcle})].

By contrast, the correspondence program for systems whose classical analogs are
chaotic has been left largely undeveloped. The nature of the classical
Liouville eigenfunctions and eigenspectrum is not clearly understood
\cite {carvalho},
and only the correspondence limits of the stationary Liouville
eigenfunctions have been explored. Berry\cite{berry2} and
Voros\cite{voros} were able to show that the Wigner function
associated with a stationary Liouville eigenstate $|n\rangle\langle
n|$ (here $r=r'=1$) concentrates in the
classical limit to a uniform distribution over the classical energy
surface of the same energy $E_n$, i.e.,
\begin{equation}
\rho_{n,n}^w({\bf x})\rightarrow h^{s/2}\frac{\delta(E_n-H({\bf x}))}{\int
d{\bf x}^{\prime}~\delta(E_n-H({\bf x}^{\prime}))}.
\end{equation}
Investigations of the semiclassical corrections to this limit have
shown that the unstable
periodic orbits of the chaotic classical system strongly influence the
quantum eigenfunctions, resulting in accentuations of probability in
the vicinity of periodic orbits\cite{berry4,berry3}. Semiclassical
corrections have also been studied for the stationary Liouville
eigenfunctions of regular systems\cite{berry1}. However, no results
have been obtained for the correspondence of the nonstationary
Liouville eigenfunctions or eigenspectrum for the chaotic case. The
reason for this is made clear in the following paper, where we show
that the $\rho_{n,m}^w({\bf x})$, $n \neq m$, for chaotic systems
do not have classical limits. Rather, one must deal with the
classical and quantum spectral projection operators discussed below.

\section{Classical Liouville Eigenfunctions and Eigenspectrum}
\label{CLEE}

The classical Liouville eigenvalue
problem [Eq. (\ref{classeiq})] requires that one find a complete set of
 orthogonal Liouville
eigenfunctions $\rho_{\lambda,\alpha}({\bf x})$ with eigenvalues
$\lambda$, i.e., they must satisfy
 \begin{equation}
\int d{\bf x}~\rho_{\lambda',\alpha'}^*({\bf
x})\rho_{\lambda,\alpha}({\bf x})=0
\end{equation}
for $\lambda\neq\lambda '$ or $\alpha\neq \alpha '$,
and
\begin{equation}
\sum_{\lambda,\alpha}\rho_{\lambda,\alpha}^*({\bf
x}_0)\rho_{\lambda,\alpha}({\bf x}) =\delta ({\bf x}-{\bf x}_0).
\end{equation}
[Note that for notational simplicity, classical
eigenvalues in the remainder of the paper are denoted $\lambda$, rather than
$\lambda^c$].
For the case of $s$ degrees of freedom and $r$ constants of
the motion we must find simultaneous eigenstates of the
complete set of $2r$ operators [Eq. (\ref{fifteen})].

If a
distribution $f({\bf x})$ is an eigenstate of a real multiplicative
operator $A({\bf x})$ with eigenvalue $A'$, then
$[A({\bf x})-A']f({\bf x})=0$.  That is, $f({\bf x})=0$
for all ${\bf x}$ not satisfying $A({\bf x})=A'$ so 
that all nontrivial solutions $f({\bf x})$ 
must be proportional to
$\delta (A({\bf x})-A')$.  That is, all distributions which are eigenstates of
the multiplicative operators $H({\bf x}),K_1({\bf
x}),\dots,K_{r-1}({\bf x})$ must be proportional to 
\begin{equation}
\delta (E-H({\bf x}))\Pi_{j=1}^{r-1}\delta(K_j({\bf x})-K_j').
\end{equation}

Little more can be said about the general case of arbitrary
$r$.  We therefore focus on the two interesting special cases:
$r=s$, corresponding to integrable systems, and $r=1$ which includes
the case of chaotic motion.

\subsection{Integrable Systems}
\label{IS1}

For $r=s$ we can introduce action
variables ${\bf I}$ and conjugate angle variables $\mbox{\boldmath
$\theta$}$  such that
$H({\bf x})=H({\bf I})$, and $K_j({\bf x})=K_j({\bf I})$ for
$j=1,\dots,s-1$. It follows that there exists an ${\bf I}'$ such that
\begin{equation}
\delta (E-H({\bf x}))\Pi_{j=1}^{r-1}\delta(K_j({\bf x})-K_j')\propto
\delta ({\bf I}({\bf x})-{\bf I}')
\label{twentyseven}
\end{equation}
and that the classical Liouville eigendistributions $\rho_{\lambda, \alpha}
(\bf x)$ must be of the form $\delta
({\bf I}({\bf x})-{\bf I}') F(\mbox{\boldmath $\theta$}({\bf x}))$.
Since
$L_c=-i\mbox{\boldmath $\omega$}({\bf I})\cdot
\frac{\partial }{\partial \mbox{\boldmath $\theta$}}$ the
Liouville eigenequation is 
\begin{equation}
-i\mbox{\boldmath $\omega$}({\bf I}')\cdot
\frac{\partial }{\partial \mbox{\boldmath $\theta$}}F(\mbox{\boldmath
$\theta$})=\lambda F(\mbox{\boldmath $\theta$})
\end{equation}
with solutions $F(\mbox{\boldmath $\theta$})=e^{i{\bf
a}\cdot\mbox{\boldmath $\theta$}}$, 
$\lambda={\bf a}\cdot\mbox{\boldmath $\omega$}({\bf I}')$. If the
system is periodic in the angle variables then ${\bf a}=
{\bf k}\in {\bf Z}^s$, i.e., ${\bf a}$ is a vector of $s$ integers.
Thus the classical Liouville eigenfunctions
for an integrable system are given by
\begin{equation}
\rho_{{\bf I}',{\bf k}}({\bf
x})=\frac{1}{(2\pi)^{s/2}}\delta({\bf I}'-{\bf I})e^{i{\bf
k}\cdot\mbox{\boldmath $\theta$}},
\label{cefs}
\end{equation}
with eigenvalue
\begin{equation}
\lambda_{{\bf I}',{\bf k}}={\bf k}\cdot\mbox{\boldmath $\omega$}({\bf I}').  
\label{rcle}
\end{equation}
Note that these eigenfunctions are also
eigenfunctions of the differential operators $i\{K_j({\bf
x}),\cdot\}=-i\frac{\partial K_j({\bf I})}{\partial {\bf I}}\cdot
\frac{\partial }{\partial \mbox{\boldmath $\theta$}}$ with
corresponding eigenvalues ${\bf k}\cdot\frac{\partial K_j({\bf
I}')}{\partial {\bf I}'}$.
If the $K_j({\bf I})=I_j$ then the eigenvalues are simply $k_j$.
This set of eigenfunctions is
complete and orthogonal, i.e.,
\begin{equation}
\sum_{\bf k}\int d{\bf I}'~\rho_{{\bf I}',{\bf k}}^*({\bf
x}')\rho_{{\bf I}',{\bf k}}({\bf x})= \delta ({\bf x}-{\bf x}'),
\end{equation}
and
\begin{equation}
\int d{\bf x}~ \rho_{{\bf I}'',{\bf k}'}^*({\bf x})\rho_{{\bf I}',{\bf
k}}({\bf x}) =\delta_{{\bf k},{\bf k}'}\delta ({\bf I}'-{\bf I}'').
\end{equation}

We comment briefly on the character of the Liouville eigenfunctions
for integrable systems; a detailed discussion and related computations
are provided elsewhere \cite{jaffe1}.
The spectrum of an integrable system on a given torus ${\bf I}'={\bf
I}({\bf x})$ is discrete and is given by the set of frequencies
$\{{\bf k}\cdot \mbox{\boldmath $\omega$}({\bf I}')|{\bf k}\in {\bf
Z}^s\}$. On the energy surface $E=H({\bf x})$, the
spectrum of an integrable system is the union of all sets $\{{\bf k}\cdot
 \mbox{\boldmath $\omega$}({\bf I}')|{\bf k}\in {\bf
Z}^s\}$ from tori ${\bf I}'={\bf I}({\bf x})$ with $H({\bf I}')=E$.
Typically, $\mbox{\boldmath $\omega$}({\bf I}')$ varies with ${\bf
I}'$, so that the spectrum of an integrable system on an energy surface is
continuous. In our terminology\cite{spec}
eigenfunctions with frequencies $\lambda_{{\bf I}',{\bf k}}={\bf k}
\cdot\mbox{\boldmath
$\omega$}({\bf I}')$ with ${\bf k}\neq 0$ belong to the continuous
spectrum while those with ${\bf k}=0$ belong to the point spectrum.
Examination of Eq.
(\ref{cefs}) reveals that the
eigenfunctions $\rho_{{\bf I}',{\bf k}}({\bf
I},\mbox{\boldmath $\theta$})$ are identically zero off the torus
 ${\bf I}({\bf x})={\bf I}'$. Eigendistributions with
 ${\bf k}=0$ are stationary and uniform on the torus, and as a
set they project out the longtime limit of any given initial
probability distribution. For ${\bf k}\neq 0$ the character of the
eigendistribution is determined by the underlying orbit structure of
the torus ${\bf I}({\bf x})={\bf I}'$. If the frequencies
$\omega_1,\omega_2,\dots,\omega_s$ of the torus are commensurate then
the orbits of
the torus ${\bf I}({\bf x})={\bf I}'$ are all periodic
\cite{footnote2}. The
eigendistributions associated with commensurate tori are nonuniform
and stationary when ${\bf k}\cdot\mbox{\boldmath $\omega$}({\bf I}')=0$
 and ${\bf k}\neq
0$, and nonuniform and nonstationary when ${\bf k}\cdot
\mbox{\boldmath $\omega$}({\bf I}')\neq 0$.

  The eigendistributions
associated with incommensurate tori are nonuniform and nonstationary
for ${\bf k}\neq 0$. Eigendistributions for integrable systems are
supported only over a given torus ${\bf I}'={\bf I}({\bf x})$, but
every orbit of the torus contributes to the construction of the
eigenfunction\cite{footnote3}. 

\subsection{Chaotic Systems}
\label{CS1}
Now consider the case where $r=1$. In addition to being chaotic we
assume that the system is hyperbolic,
i.e., that it has a countable number of isolated unstable periodic orbits on
each surface of constant energy. Thus, our chaotic system has
positive Kolmogorov entropy, exhibits sensitivity to initial
conditions (i.e., possesses $2s-2$ nonvanishing Liapunov exponents), and
is mixing, and hence is
weak mixing and ergodic\cite{arnold}. Such familiar properties of
chaotic systems (i.e., ergodicity, weak
mixing, mixing etc.) may be related to the spectral properties of the
Liouville operator\cite{arnold}. For example, if $\lambda=0$ is a
nondegenerate point eigenvalue of the Liouville operator $L_c$ on the
energy surface then the system is ergodic. If $\lambda=0$ is the only
point eigenvalue of $L_c$ on the energy surface then the system
is weak mixing. If the rest of the spectrum is continuous then the
system is mixing. In this section we determine the form of the
generalized  eigenfunctions which correspond to the point and continuous
Liouville eigenspectrum\cite{arnold,spec} for chaotic systems. 

In accord with Eq. (\ref{fifteen}) we seek eigenfunctions of the
complete set $L_c$ and
$H(\bf x)$. As
eigenstates of the energy multiplication operator $H({\bf
x})$ they must be proportional to $\delta(E-H({\bf
x}))$ where $E$ is the corresponding energy eigenvalue. Consider then
the
function $\tau ({\bf x})$ which is
conjugate to the constant of the motion $H({\bf x})$, i.e., $\{\tau
({\bf x}),H({\bf x})\}=1$.  Then
the distribution $\delta(E-H({\bf x}))e^{i\lambda \tau({\bf x})}$
is an eigenfunction of the classical Liouville operator with
eigenvalue $\lambda$ since $L_c=i\{H({\bf x}),\cdot\}=-i\frac{\partial}{\partial
\tau}$.  Note that $e^{-iL_ct}\tau({\bf x})=\tau ({\bf X}({\bf
x},-t))=\tau({\bf x})-t$, where ${\bf X}({\bf x},-t)$ is the phase
space point from which ${\bf x}$ evolves over the time $t$. Thus,
for a given choice of the origin of time along each orbit of the
system, we can construct a function $\tau ({\bf x})$ which
specifies the time of an arbitrary point ${\bf x}$ along the orbit,
since trajectories do not intersect one another in phase space.
However, the distributions $\delta(E-H({\bf
x}))e^{i\lambda \tau({\bf x})}$ \cite{footnote4}, do not form a complete 
set of Liouville eigenfunctions since they do not account for the infinite
degeneracy of the classical Liouville spectrum for chaotic
systems\cite{arnold}. 

To properly define eigenfunctions we must consider the manner in which
points in phase space are labeled. Specifically we introduce a
new system of variables: $H({\bf x})$  which is the energy of a point
in phase space, $\tau({\bf x})$ a function which assigns a time to every
point in phase space, and $2s-2$ stationary variables $\mbox{\boldmath
$\eta$}({\bf x})$ which label the
trajectories on the energy surface. The functions $\tau$ and
$\mbox{\boldmath $\eta$}$ are continuous
only along an orbit: $\tau$ increments smoothly while $\mbox{\boldmath
$\eta$}$ remains constant along an orbit.
The $\mbox{\boldmath $\eta$}$ variables are chosen to have the property that
\begin{equation}
\delta({\bf x}_0-{\bf x})=\delta(H({\bf
x}_0)-H({\bf
x}))\delta(\tau({\bf
x}_0)-\tau({\bf
x}))\delta(\mbox{\boldmath $\eta$}({\bf
x}_0)-\mbox{\boldmath $\eta$}({\bf
x})).
\label{le1}
\end{equation}
 As one moves off a
trajectory the variables $\tau$ and $\mbox{\boldmath $\eta$}$ must take on all
possible values infinitely often since (a) there are an infinite number of chaotic
trajectories, (b) every chaotic trajectory comes arbitrarily
close to every phase space point on every other trajectory, and (c) chaotic
trajectories return arbitrarily close to themselves infinitely often.
We now introduce a complete set of square integrable orthonormal functions
 $\chi_\ell(\mbox{\boldmath $\eta$})$, $\ell\in{\bf Z}$, i.e.,
\begin{equation}
\sum_\ell \chi_\ell^*(\mbox{\boldmath $\eta$}({\bf
x}_0))\chi_\ell(\mbox{\boldmath $\eta$}({\bf
x}))=\delta(\mbox{\boldmath $\eta$}({\bf
x}_0)-\mbox{\boldmath $\eta$}({\bf
x}))
\label{le2}
\end{equation}
and
\begin{equation}
\int d\mbox{\boldmath $\eta$}~\chi_{\ell'}^*(\mbox{\boldmath
$\eta$})\chi_\ell(\mbox{\boldmath $\eta$})=\delta_{\ell',\ell}.
\label{le3}
\end{equation}
Classical Liouville eigenfunctions for chaotic systems
$\rho_{E,\lambda}^\ell ({\bf x})$
are then defined as:
\begin{equation}
\rho_{E,\lambda}^\ell ({\bf x})=\frac{1}{\sqrt{2\pi}}\delta(E-H({\bf
x}))e^{i\lambda\tau({\bf x})}\chi_\ell(\mbox{\boldmath $\eta$}({\bf x})).
\label{grail}
\end{equation}
Here the integer $\ell$ labels the infinite 
but countable degeneracy of the Liouville eigenvalue
$\lambda$ for the continuous part of the spectrum\cite{arnold} at energy $E$.
The distributions $\rho_{E,\lambda}^\ell ({\bf x})$ are supported over
the entire energy surface $E=H({\bf x})$ and
so are in some sense constructed from every orbit of the energy
surface.  
It can be readily verified that these eigenfunctions are both
orthogonal and complete (see Appendix A):
\begin{eqnarray}
\int d{\bf x} ~\rho_{E',\lambda'}^{\ell'*} ({\bf
x})\rho_{E,\lambda}^\ell
({\bf x}) &=&
\delta_{\ell,\ell'}\delta(E-E')\delta(\lambda-\lambda'),\nonumber\\
\sum_\ell\int_0^{\infty}dE\int_{-\infty}^{\infty}
d\lambda~\rho_{E,\lambda}^{\ell*} ({\bf x}_0)\rho_{E,\lambda}^\ell ({\bf
x}) &=& \delta({\bf x}-{\bf x}_0).
\label{sdec}
\end{eqnarray}
Here we note that the introduction of the complete set of $\chi_\ell$ is
necessary to achieve the correct $\delta(\bf x-\bf x_0)$ term in Eq.
(\ref{sdec}) and that these eigenfunctions are quite different from
those proposed elsewhere which we have shown to be incorrect \cite{carvalho}.

The eigenfunctions in Eq. (\ref{grail}) have a structure which
at first sight appears to violate several well known properties
of chaotic systems. For example, the existence of stationary
square integrable functions $\chi_\ell(\mbox{\boldmath $\eta$}({\bf
x}))$, which are not functions of the Hamiltonian, appears to contradict the fact
that for ergodic systems the only stationary square integrable
distributions are functions of the Hamiltonian. However, the
$\chi_\ell(\mbox{\boldmath $\eta$}({\bf x}))$ distributions are not square
integrable on the energy surface. That is,
\begin{equation}
\int dHd\tau d\mbox{\boldmath $\eta$}~\delta(E-H)~|\chi_
\ell(\mbox{\boldmath $\eta$}({\bf x}))|^2=\int
d\tau=\infty .
\end{equation}
In addition, it might appear objectionable that $\chi_\ell(\mbox{\boldmath
$\eta$}({\bf x}))$
is a global constant of the motion
independent of the Hamiltonian. However, this is not the case since the
set of points ${\bf x}$ satisfying
$\chi_\ell(\mbox{\boldmath $\eta$}({\bf x}))=\chi'$, where $\chi'$ is some
constant, at most contains points from a countable number of trajectories
and so cannot divide the phase space in any meaningful way. Thus the
functions $\chi_\ell(\mbox{\boldmath $\eta$}({\bf x}))$ are not true
constants of the motion.

The relation between these eigenfunctions and the Liouville spectrum
on the energy surface $E=H({\bf x})$ is not immediately clear.
Examination of
the spectral decomposition in Eq. (\ref{sdec}) reveals no
separation of point and continuous spectrum, although the Liouville
eigenfunctions have the correct degeneracy for the continuous
spectrum. A proper treatment of the separation of point and continuous
spectrum is given in Sec. \ref{CS2}.

\section{Correspondence: Integrable Systems}
\label{corres}

Having established the essential features of the classical eigenstate
picture we now examine correspondence. We assume that $r'=r$, i.e., that
the complete set of commuting quantum observables has as many members as
there are constants of the classical motion.

Given the complete
set of commuting observables  $\hat{H},\hat{K}_1,\dots,\hat{K}_{r-1}$ for
$r\leq s$,
the complete set of $2r$ commuting superoperators is given by [Eq.
(\ref{Lcal})]:
\begin{equation}
\hat{L},\hat{{\cal
H}},\frac{1}{\hbar}[\hat{K}_1,\cdot],\frac{1}{2}
[\hat{K}_1,\cdot]_+,\dots,\frac{1}{\hbar}[\hat{K}_{r-1},\cdot],
\frac{1}{2}[\hat{K}_{r-1},\cdot]_+.
\end{equation}
We denote the simultaneous eigenstates of
the $r$ observables by $|{\bf n}\rangle$ where ${\bf
n}=(n_1,\dots,n_r)$ are the quantum numbers, and the corresponding
eigenvalues are $E_{{\bf n}},K_1({\bf n}),\dots,K_{r-1}({\bf
n})$. We further denote the Wigner-Weyl representation of the quantum
observables $\hat{H}, \hat{K}_j$ by $H({\bf x})$, and $K_j^w({\bf x})$,
for $j=1,\dots,r-1$.

The Liouville eigenstates $\rho_{{\bf n},{\bf
m}}^w$ must be an eigenstate of a corresponding set of $2r$ super-operators in
the Wigner-Weyl representation which are 
\begin{eqnarray}
&&L({\bf x}),{\cal H}({\bf x}),\frac{2i}{\hbar}K_j^w({\bf x})\sin
(\frac{\hbar\sigma}{2}),K_j^w({\bf x})\cos
(\frac{\hbar\sigma}{2}),\dots\nonumber \\
&&~~~~~~~~\dots, \frac{2i}{\hbar}K_{r-1}^w({\bf x})\sin
(\frac{\hbar\sigma}{2}),K_{r-1}^w({\bf x})\cos
(\frac{\hbar\sigma}{2}),
\label{qset}
\end{eqnarray}
with 
$L({\bf x})=(2i/\hbar)H({\bf x})\sin
(\hbar\sigma/2)$ and ${\cal H}({\bf x})= H({\bf x})\cos
(\hbar\sigma/2)$. 
By comparison the classical complete set is
\begin{equation}
L_c({\bf x}),H({\bf x}), i\{K_1({\bf x}),\cdot\},K_1({\bf x}), \dots,
i\{K_{r-1}({\bf x}),\cdot\},K_{r-1}({\bf x}).
\label{cset}
\end{equation}
We assume that $\lim_{h\rightarrow 0} K_j^w({\bf x})=K_j(x)$. Since 
$\lim_{h\rightarrow 0}\cos(\hbar\sigma/2)=1$ and $\lim_{h\rightarrow 0}
\left({\frac{2i}{\hbar}} \right) \sin(\hbar\sigma/2)=i\sigma$, we note that
formally the
quantum operator set [Eq. (\ref{qset})] becomes the classical operator
set [Eq. (\ref{cset})] in the classical limit.  The first eigenequation is the time
independent von Neumann (quantum Liouville) equation
\begin{equation}
L({\bf x})\rho_{{\bf n},{\bf m}}^w({\bf x})=\lambda_{{\bf n},{\bf m}}\rho_{{\bf
n},{\bf m}}^w({\bf x}), 
\label{qliouv}
\end{equation}
where $L({\bf x})$ is the quantum Liouville operator in the
Wigner-Weyl representation
and $\lambda_{{\bf n},{\bf m}}=(E_{{\bf n}}-E_{{\bf m}})/\hbar$.
Second,
$\rho_{{\bf n},{\bf m}}^w$ must satisfy the eigenequation 
for the Hermitian energy operator $\hat{{\cal H}}
=\frac{1}{2}[\hat{H},~]_+$, i.e., the anti-commutator equation
\begin{equation}
\frac{1}{2}([\hat{H},|{\bf n}\rangle\langle {\bf m}|]_{+})^w \equiv H({\bf
x})\cos(\hbar\sigma/2)\rho_{{\bf n},{\bf m}}^w({\bf
x}) =\frac{E_{{\bf n}}+E_{{\bf m}}}{2}\rho_{{\bf n},{\bf m}}^w({\bf x}),
\end{equation}
or 
\begin{equation}
{\cal H}({\bf x})\rho_{{\bf n},{\bf m}}^w({\bf x})=E_{{\bf n},{\bf
m}}\rho_{{\bf n},{\bf m}}^w({\bf x}),
\label{qheig}
\end{equation}
where $E_{{\bf n},{\bf m}}=(E_{{\bf n}}+E_{{\bf m}})/2$.
Similarly, for $j=1,\dots,r-1$, eigenfunctions $\rho_{{\bf n},{\bf
m}}^w({\bf x})$ satisfy the equations
\begin{equation}
(\frac{1}{\hbar}[\hat{K}_j,|{\bf n}\rangle\langle {\bf m}|])^w \equiv \frac{2i}
{\hbar}K_j^w({\bf x})\sin(\hbar\sigma/2)\rho_{{\bf n},{\bf m}}^w({\bf x})
 =\frac{K_j({\bf n})- K_j({\bf m})}{\hbar}\rho_{{\bf n},{\bf m}}^w({\bf x}),
\label{k-}
\end{equation}
and
\begin{equation}
(\frac{1}{2}[\hat{K}_j,|{\bf n}\rangle\langle
 {\bf m}|]_{+})^w \equiv K_j^w({\bf x})\cos(\hbar\sigma/2)
  \rho_{{\bf n},{\bf m}}^w({\bf x})
=\frac{K_j({\bf n})+ K_j({\bf m})}{2}\rho_{{\bf n},{\bf m}}^w({\bf x}).
\label{k+}
\end{equation}

We can readily explore the classical limit of the $2r$ quantum eigenequations,
Eqs. (\ref{qliouv})-(\ref{k+}) if 
\begin{equation}
\lim_{h \rightarrow 0}
\rho_{{\bf n},{\bf m}}^w({\bf
x})\rightarrow \rho_{{\bf n},{\bf m}}^c({\bf x}),
\label{rhol}
\end{equation}
where $\rho_{{\bf n},{\bf m}}^c({\bf x})$ is to be determined.  
However, this assumption is not generally valid. That is,
for quantum systems whose classical
analogs are chaotic ($r=r'=1$), as we demonstrate in Ref. \cite{chacor},
individual quantum eigenfunctions
$\rho_{n,m}^w({\bf x})$, $n\neq m$, display essential singularities
as $h\rightarrow 0$, and hence do not have correspondence limits. 
Rather, the spectral
projection operators introduced later below [Sect. V] do have 
correspondence limits. Nonetheless,  we can adopt 
Eq. (\ref{rhol}) primarily for notational convenience; the
arguments which follow can be reformulated in terms of quantum and
classical spectral projection operators when Eq. (\ref{rhol}) is invalid.

We assume that
the $h \rightarrow 0$ limit of $K_j^w({\bf x})$ is $K({\bf x})$.
In particular, if we consider the case where the set of $2r$ quantum operators 
[Eq. (\ref{qset})] gives the classical set [Eq. (\ref{cset})] in the
classical limit and that Eq. (\ref{rhol}) holds then each of the
eigenvalue equations [Eqs. (\ref{qliouv})-(\ref{k+})] reduces to the
set of classical eigenvalue equations for $\rho_{\lambda, \alpha}({\bf
x})$ with the same eigenvalues as in the quantum eigenequations. If this is
the case then 
the quantum $\rho^w_{\bf n,m}({\bf x})$ goes to the classical
$\rho_{\lambda, \alpha}({\bf x})$, with clearly identifiable eigenvalues and
correspondence is established.

An example of this approach, applied to integrable systems, is provided
below.

\subsection{Integrable Systems:  Formal Correspondence}
\label{AIS}

In this section we first apply this approach to determine the
complete set of correspondence rules for integrable systems. 
The resulting picture of
correspondence for integrable systems is exceptionally clear but
gives no insight into the approach to the classical limit. This is
treated in the next subsection.
 
In this instance Eq. (\ref{rhol}) has been proven by Berry\cite{berry1} and by
Jaff\'{e} and Brumer\cite{jaffe2}. Thus, in accord with the previous section,
the eigenfunctions $\rho_{{\bf n},{\bf m}}^w({\bf
x})$ satisfy $2s$ classical eigenequations in the classical limit:
\begin{equation}
L_c({\bf x})\rho_{{\bf n},{\bf m}}^c({\bf x})=\lambda_{{\bf n},{\bf m}}
\rho_{{\bf n},{\bf m}}^c({\bf x}), 
\label{c1}
\end{equation}
\begin{equation}
H({\bf x})\rho_{{\bf n},{\bf m}}^c({\bf x})=E_{{\bf n},{\bf
m}}\rho_{{\bf n},{\bf m}}^c({\bf x}),
\label{c2}
\end{equation}
\begin{equation}
i\{K_j({\bf
I}),\cdot\}\rho_{{\bf n},{\bf m}}^c({\bf
I},\mbox{\boldmath $\theta$})=\frac{K_j({\bf n})- K_j({\bf m})}{\hbar}
\rho_{{\bf n},{\bf m}}^c({\bf
I},\mbox{\boldmath $\theta$})=0,
\label{c3}
\end{equation}
\begin{equation}
K_j({\bf
I})\rho_{{\bf n},{\bf m}}^c({\bf
I},\mbox{\boldmath $\theta$})=\frac{K_j({\bf n})+ K_j({\bf m})}{2}
\rho_{{\bf n},{\bf m}}^c({\bf
I},\mbox{\boldmath $\theta$})=0,
\label{c4}
\end{equation}
for $j=1,\dots, s-1$. As in Eq. (\ref{twentyseven})  
we conclude that $\rho_{{\bf n},{\bf m}}^c({\bf x})$ is
proportional to
\begin{equation}
\delta (E-H({\bf x}))\Pi_{j=1}^{s-1}\delta(K_j({\bf x})-K_j'),
\end{equation}
where $E=E_{{\bf n},{\bf m}}$, and $K_j'=\frac{K_j({\bf n})+K_j({\bf
m})}{2}$. Since $H({\bf x})=H({\bf I})$, and $K_j({\bf x})=K_j({\bf
I})$ for $j=1,\dots,s-1$ there exists an ${\bf I}_{{\bf
n},{\bf m}}$ such that
\begin{equation}
\delta (E-H({\bf x}))\Pi_{j=1}^{s-1}\delta(K_j({\bf
x})-K_j')\propto \delta({\bf I}-{\bf I}_{{\bf
n},{\bf m}}).
\end{equation}
Hence $\rho_{{\bf n},{\bf m}}^c({\bf x})\propto \delta({\bf I}-{\bf I}_{{\bf
n},{\bf m}}) F_{{\bf n},{\bf m}}(\mbox{\boldmath $\theta$})$.

>From Eqs. (\ref{c1}) and (\ref{c3}) it then follows that
\begin{equation}
-i\mbox{\boldmath $\omega$}({\bf I}_{{\bf n},{\bf m}})\cdot
\frac{\partial }{\partial \mbox{\boldmath $\theta$}}F_{{\bf n},{\bf m}}
(\mbox{\boldmath
$\theta$})=\frac{E_{{\bf n}}-E_{{\bf m}}}{\hbar} F_{{\bf n},{\bf
m}}(\mbox{\boldmath $\theta$})
\label{c5}
\end{equation}
and
\begin{equation}
-i\frac{\partial K_j}{\partial {\bf I}}({\bf I}_{{\bf n},{\bf m}})\cdot
\frac{\partial }{\partial \mbox{\boldmath $\theta$}}F_{{\bf n},{\bf m}}
(\mbox{\boldmath
$\theta$})=\frac{K_j({\bf n})-K_j({\bf m})}{\hbar} F_{{\bf n},{\bf
m}}(\mbox{\boldmath $\theta$}).
\label{c6}
\end{equation}
By WKB \cite{jaffe2,berry1,delos} $E_{{\bf n}}\sim H({\bf I}_{\bf n})$
where ${\bf I}_{\bf n}=({\bf n}+\mbox{\boldmath $\beta$})\hbar$ and
where
$\mbox{\boldmath $\beta$}$ are the Maslov indices. Similarly we may
approximate $K_j({\bf n})\sim K_j({\bf I}_{\bf n})$. Assuming that
\begin{equation}
{\bf I}_{{\bf n},{\bf m}}-{\bf I}_{\bf n}\sim O(h)\sim {\bf I}_{{\bf
n},{\bf m}}-{\bf I}_{\bf m} <<{\bf I}_{{\bf n},{\bf m}}
\label{small}
\end{equation}
it follows that
\begin{eqnarray}
\frac{E_{{\bf n}}-E_{{\bf m}}}{\hbar}&\sim& \frac{H({\bf I}_{\bf
n})-H({\bf I}_{{\bf m}})}{\hbar}\nonumber \\
&\sim& \frac{H({\bf I}_{{\bf n},{\bf m}}-({\bf I}_{{\bf n},
{\bf m}}-{\bf I}_{\bf
n}))-H({\bf I}_{{\bf n},{\bf m}}-({\bf I}_{{\bf n},{\bf m}}-{\bf
I}_{{\bf m}})}{\hbar}\nonumber \\
&\sim &\mbox{\boldmath $\omega$}({\bf I}_{{\bf n},{\bf
m}})\cdot ({\bf n}-{\bf m})
\label{c7}
\end{eqnarray}
by the Taylor expansion of $H({\bf I}_{\bf n})$ and $H({\bf I}_{\bf m})$
about ${\bf I}_{{\bf n},{\bf m}}$.
Similarly one can show that
\begin{equation}
\frac{K_j({\bf n})-K_j({\bf m})}{\hbar}\sim \frac{\partial K_j}{\partial
 {\bf I}}({\bf I}_{{\bf n},{\bf
m}})\cdot ({\bf n}-{\bf m}).
\label{c8}
\end{equation}
Substituting Eqs. (\ref{c7}) and (\ref{c8}) into Eqs.
(\ref{c5}) and (\ref{c6}) shows that $F_{{\bf n},{\bf
m}}(\mbox{\boldmath $\theta$})\sim e^{i({\bf n}-{\bf
m})\cdot\mbox{\boldmath $\theta$}}$.  Thus, up to a
normalization factor
\begin{equation}
\rho_{{\bf n},{\bf m}}^w({\bf x})\rightarrow
\hbar^{s/2}\rho_{{\bf I}_{{\bf n},{\bf
m}},{\bf n}-{\bf m}}({\bf x})
\label{gen}
\end{equation}
in the limit as $h\rightarrow 0$. 

More specifically, we have to see that
$\rho_{{\bf n},{\bf m}}^w$, an eigenfunction of the $2s$
quantum operators 
\begin{eqnarray}
&&L({\bf x}),{\cal H}({\bf x}),\frac{2i}{\hbar}K_j^w({\bf x})\sin
(\frac{\hbar\sigma}{2}),K_j^w({\bf x})\cos
(\frac{\hbar\sigma}{2}),\dots\nonumber \\
&&~~~~~~~~\dots, \frac{2i}{\hbar}K_{s-1}^w({\bf x})\sin
(\frac{\hbar\sigma}{2}),K_{s-1}^w({\bf x})\cos
(\frac{\hbar\sigma}{2}),
\end{eqnarray}
with corresponding eigenvalues
\begin{eqnarray}
&&\lambda_{{\bf n},{\bf m}},E_{{\bf n},{\bf m}},\frac{K_1({\bf
n})-K_1({\bf m})}{\hbar},\frac{K_1({\bf
n})+K_1({\bf m})}{2},\dots\nonumber \\
&&~~~~~~~~\dots,\frac{K_{s-1}({\bf
n})-K_{s-1}({\bf m})}{\hbar},\frac{K_{s-1}({\bf
n})+K_{s-1}({\bf m})}{2}
\end{eqnarray}
approaches a classical limit $\rho_{{\bf n},{\bf m}}^c({\bf I},\mbox{\boldmath
$\theta$})=\hbar^{s/2}\rho_{{\bf I}_{{\bf n},{\bf
m}},{\bf n}-{\bf m}}({\bf x})$ which is an eigenfunction of the $2s$
classical operators
\begin{equation}
L_c({\bf x}),H({\bf x}), i\{K_1({\bf x}),\cdot\},K_1({\bf x}), \dots,
i\{K_{s-1}({\bf x}),\cdot\},K_{s-1}({\bf x}),
\end{equation}
with corresponding eigenvalues
\begin{eqnarray}
&&\lambda_{{\bf I}_{{\bf n},{\bf m}},{\bf n}-{\bf m}},E_{{\bf n},{\bf
m}},({\bf n}-{\bf m})\cdot \frac{\partial K_1}{\partial {\bf I}}({\bf
I}_{{\bf n},{\bf m}}),\frac{K_1({\bf n})+K_1({\bf
m})}{2},\dots\nonumber \\
&&~~~~~~~~\dots,({\bf
n}-{\bf m})\cdot \frac{\partial K_{s-1}}{\partial {\bf I}}({\bf
I}_{{\bf n},{\bf m}}),\frac{K_{s-1}({\bf n})+K_{s-1}({\bf m})}{2}.
\end{eqnarray}

Using a completely different argument which employed primitive WKB
wavefunctions Berry showed \cite{berry1} that the stationary
Liouville eigenfunctions have the correspondence limit
\begin{equation}
\rho_{{\bf n},{\bf n}}^w({\bf x})\rightarrow
\hbar^{s/2}\rho_{{\bf I}_{{\bf n}},0}({\bf x}),
\end{equation}
where ${\bf I}_{{\bf n}}=[{\bf n}+\mbox{\boldmath $\beta$}]\hbar$.
Since ${\bf I}_{{\bf n},{\bf n}}\equiv {\bf I}_{\bf n}$ both results are
in agreement.
For the case of nonstationary Liouville eigenfunctions Jaff\'{e}
and Brumer \cite{jaffe2} proved the general formula, Eq. (\ref{gen}),
where they surmised that ${\bf I}_{{\bf n},{\bf m}}=({\bf I}_{\bf
n}+{\bf I}_{\bf m})/2$. However, ${\bf I}_{{\bf n},{\bf m}}$
is obtained correctly by solving the $s$ equations
\begin{equation}
(E_{{\bf n}}+E_{{\bf m}})/2=H({\bf I}_{{\bf
n},{\bf m}}), 
\label{eee}
\end{equation}
and
\begin{equation}
(K_j({\bf n})+ K_j({\bf m}))/2=K_j({\bf
I}_{{\bf
n},{\bf m}})
\label{jjj}
\end{equation}
for $j=1,\dots, s-1$.
The surmise of Jaff\'{e} and Brumer\cite{jaffe2} is an approximation
which is good to first order in $h$. To see this note that for
$\mbox{\boldmath $\omega$}(0)\neq 0$ we can expand
$H({\bf I})\sim \mbox{\boldmath $\omega$}(0)\cdot{\bf I} +O({\bf I}^2)$.
Equation (\ref{eee}) then reduces to $\mbox{\boldmath
$\omega$}(0)\cdot[{\bf I}_{{\bf n},{\bf m}}-({\bf I}_{{\bf n}}+{\bf
I}_{{\bf m}})/2] +O(h^2)\sim 0$ for which ${\bf I}_{{\bf n},{\bf m}}
\sim({\bf I}_{{\bf n}}+{\bf
I}_{{\bf m}})/2$ is clearly a solution to first order in $h$. 
If $\partial K_j(0)/
\partial {\bf I}\neq 0$ for all $j$ then it can be
shown in a similar fashion that ${\bf I}_{{\bf n},{\bf m}}
\sim({\bf I}_{{\bf n}}+{\bf
I}_{{\bf m}})/2$ is a first order solution of Eqs. (\ref{jjj}). Thus,
the approximation \cite{jaffe2} ${\bf I}_{{\bf
n},{\bf m}}\sim ({\bf I}_{{\bf n}}+{\bf
I}_{{\bf m}})/2$ is good to first order in Planck's constant. 

\subsection{Integrable Systems:  Approach to the Limit}
\label{1WKB}

The method described above assumes the validity of Eq. (\ref{rhol})
and fails to provide insights into the behavior of the
eigendistributions as $h\rightarrow 0$. Here we extend and reexamine
our previous proof\cite{jaffe2} of Eq. (\ref{rhol}) for integrable systems with
particular interest in the mechanism by which essential singularities
are avoided as $h\rightarrow 0$.
The following argument also determines the correspondence limits of the
stationary and nonstationary Liouville eigenfunctions in a unified manner.

In the $h\rightarrow 0$ limit the eigenfunctions of the Hamiltonian,
$\langle {\bf q}|{\bf n}\rangle$, are well approximated by linear
combinations of primitive WKB wavefunctions, i.e.,
\begin{equation}
\langle {\bf q}|{\bf n}\rangle\sim \sum_j(2\pi)^{-s/4}|{\rm det}
\frac{\partial^2R_j({\bf q},{\bf I}_{\bf n})}{\partial {\bf
q}\partial {\bf I}}|^{1/2}e^{iR_j({\bf q},{\bf I}_{\bf n})/\hbar}
e^{-i\nu_j\pi/2}
\end{equation}
where $R_j({\bf q},{\bf I})=\int_{{\bf q}_0}^{{\bf q}}{\bf p}_j({\bf
q}',{\bf I})\cdot d{\bf q}'$ are the actions corresponding to the
different branches of the multivalued momentum functions ${\bf
p}_j({\bf q},{\bf I})$, and $\nu_j$ are the Maslov indices. Thus, 
it follows that
\begin{equation}
\rho_{{\bf n},{\bf m}}^w({\bf x})=h^{-s/2}\int d{\bf v}e^{i{\bf p}\cdot
{\bf v}/\hbar}\langle {\bf q}-{\bf v}/2|{\bf n}\rangle\langle {\bf
m}|{\bf q}+{\bf v}/2\rangle \sim\sum_{j,k}C_{{\bf n},{\bf m}}^{j,k}({\bf x}),
\label{always}
\end{equation}
where
\begin{eqnarray}
C_{{\bf n},{\bf m}}^{j,k}({\bf x})&=&
(2\pi h)^{-s/2}\int d{\bf v}|{\rm det}\frac{\partial^2R_j({\bf
q}-{\bf v}/2,{\bf I}_{\bf n})}{\partial {\bf
q}\partial {\bf I}}{\rm det}\frac{\partial^2R_k({\bf q}+{\bf v}/2,
{\bf I}_{\bf m})}{\partial {\bf
q}\partial {\bf I}}|^{1/2}\cdot \nonumber \\
&&e^{i{\bf p}\cdot {\bf v}/\hbar}e^{iR_j({\bf q}-{\bf v}/2,{\bf I}_{\bf
n})/\hbar}e^{-iR_k({\bf q}+{\bf v}/2,{\bf I}_{\bf m})/\hbar}
e^{i(\nu_j-\nu_k)\pi/2},
\label{always2}
\end{eqnarray}
and where, in principle, we must retain all contributions to the double sum
from the different sheets of the phases. However, one of us
has argued \cite{jaffe2} that the $j\neq k$ terms should make a negligible
contribution to
$\rho_{{\bf n},{\bf m}}^w({\bf x})$ due to the rapidly oscillatory
character of their integrands. Before examining the $j=k$ terms which yield 
the classical limit as $h\rightarrow 0$, we briefly demonstrate 
that the $j\neq k$ terms are indeed small if some averaging over phase
space (essentially an energy average) is performed. (We would 
prefer to carry out an average over wavefunctions at different
energies, as in the chaotic case\cite{chacor}, but it is 
unclear as to how to do this.  We expect the phase
space average to achieve a similar result).

Making the change of variable ${\bf v}\rightarrow h{\bf v}$ Eq. (\ref{always2})
can be written in the form 
\begin{eqnarray}
C_{{\bf n},{\bf m}}^{j,k}({\bf x})&=&\hbar^{s/2}
\int d{\bf v}|{\rm det}\frac{\partial^2R_j({\bf
q}-h{\bf v}/2,{\bf I}_{\bf n})}{\partial {\bf
q}\partial {\bf I}}{\rm det}\frac{\partial^2R_k({\bf q}+h{\bf v}/
2,{\bf I}_{\bf m})}{\partial {\bf
q}\partial {\bf I}}|^{1/2}\cdot \nonumber \\
&&e^{2\pi i{\bf p}\cdot {\bf v}}e^{iR_j({\bf q}-h{\bf v}/2,{\bf I}_{\bf
n})/\hbar}e^{-iR_k({\bf q}+h{\bf v}/2,{\bf I}_{\bf m})/\hbar}
e^{i(\nu_j-\nu_k)\pi/2}.
\label{expanding}
\end{eqnarray}
Note the presence of essential singularities
in each of the factors $e^{iR({\bf q}-h{\bf v}/2,{\bf I}_{\bf
n})/\hbar}$ and $e^{-iR({\bf q}+h{\bf v}/2,{\bf I}_{\bf m})/\hbar}$.
Expanding the actions $R$ in powers of $h$, we have
\begin{equation}
R_j({\bf q}-h{\bf v}/2,{\bf I}_{\bf
n})\sim R_j({\bf q},{\bf I}_{\bf n})-\frac{h}{2}{\bf p}_j({\bf q},
{\bf I}_{\bf n})\cdot 
\tilde{{\bf v}}+\frac{h^2}{8}{\bf v}\cdot \frac{\partial {\bf p}_j({\bf q},
{\bf I}_{\bf n})}{\partial {\bf q}}\cdot \tilde{{\bf v}},
\end{equation}
and
\begin{equation}
R_k({\bf q}+h{\bf v}/2,{\bf I}_{\bf
m})\sim R_k({\bf q},{\bf I}_{\bf m})+\frac{h}{2}{\bf p}_k
({\bf q},{\bf I}_{\bf m})\cdot \tilde{{\bf v}}+\frac{h^2}{8}{\bf v}
\cdot \frac{\partial {\bf p}_k({\bf q},{\bf I}_{\bf m})}
{\partial {\bf q}}\cdot \tilde{{\bf v}}.
\end{equation}
We also assume that the determinant factors in the integrand 
are slowly varying, i.e.,
\begin{equation}
|{\rm det}\frac{\partial^2R_j({\bf
q}-h{\bf v}/2,{\bf I}_{\bf n})}{\partial {\bf
q}\partial {\bf I}}{\rm det}\frac{\partial^2R_k({\bf q}+h{\bf v}/2,
{\bf I}_{\bf m})}{\partial {\bf
q}\partial {\bf I}}|^{1/2}\sim |{\rm det}\frac{\partial^2R_j({\bf
q},{\bf I}_{\bf n})}{\partial {\bf
q}\partial {\bf I}}{\rm det}\frac{\partial^2R_k({\bf q},{\bf I}_{\bf m})}
{\partial {\bf q}\partial {\bf I}}|^{1/2}.
\end{equation}
Here the tilde denotes the transpose (a column vector) of the row vector {\bf
v}. Substituting these expressions into Eq. (\ref{expanding}) and performing
the 
integral over ${\bf v}$ shows that Eq. (\ref{expanding}) becomes
\begin{eqnarray}
&&C_{{\bf n},{\bf m}}^{j,k}({\bf x})=(2i/\pi)^{s/2}|{\rm det}\frac{\partial^2R_j({\bf
q},{\bf I}_{\bf n})}{\partial {\bf
q}\partial {\bf I}}{\rm det}\frac{\partial^2R_k({\bf q},{\bf I}_{\bf m})}
{\partial {\bf q}\partial {\bf I}}|^{1/2}[{\rm det}(\frac{\partial {\bf p}_j({\bf q},
{\bf I}_{\bf n})}{\partial {\bf q}}-\frac{\partial {\bf p}_k({\bf q},
{\bf I}_{\bf m})} {\partial {\bf q}})]^{-1/2} \cdot \nonumber \\
&&e^{i(R_j({\bf q},{\bf I}_{\bf n})-R_k({\bf q},{\bf I}_{\bf m}))/\hbar} 
\exp \{ {-2i ({\bf p}-{\bf
p}_{j,k})}\cdot [\frac{\partial {\bf p}_j({\bf q},
{\bf I}_{\bf n})}{\partial {\bf q}}-\frac{\partial 
{\bf p}_k({\bf q},{\bf I}_{\bf m})}{\partial {\bf q}}]^{-1}\cdot 
\widetilde{({\bf p}-{\bf p}_{j,k})}/\hbar \}e^{i(\nu_j-\nu_k)\pi/2}
\label{unsmo}
\end{eqnarray}
where ${\bf p}_{j,k}=({\bf p}_j({\bf q},{\bf I}_{\bf n})+
{\bf p}_k({\bf q},{\bf I}_{\bf m}))/2$. The contributions $C_{{\bf n},{\bf m}}^{j,k}({\bf x})$
are therefore highly oscillatory functions of the phase space variables. 

We now smooth Eq. (\ref{unsmo}) over intervals of length $\Delta p$ about each
momentum variable $p_l$, $l=1,\dots s$. In the stationary phase approximation 
(for $h\rightarrow 0$) 
\begin{eqnarray}
&&(\Delta p)^{-s}\int_{-\Delta p/2}^{\Delta p/2}d{\bf u}~[{\rm det}(\frac{\partial {\bf p}_j({\bf q},
{\bf I}_{\bf n})}{\partial {\bf q}}-\frac{\partial {\bf p}_k({\bf q},
{\bf I}_{\bf m})} {\partial {\bf q}})]^{-1/2}\cdot \nonumber \\
&&\exp \{ {-2i ({\bf p}+{\bf u}-{\bf
p}_{j,k})}\cdot [\frac{\partial {\bf p}_j({\bf q},
{\bf I}_{\bf n})}{\partial {\bf q}}-\frac{\partial 
{\bf p}_k({\bf q},{\bf I}_{\bf m})}{\partial {\bf q}}]^{-1}\cdot 
\widetilde{({\bf p}+{\bf u}-{\bf p}_{j,k})}/\hbar \}\nonumber \\
&&\rightarrow (\pi\hbar/2i)^{s/2}\Omega_{\Delta p}({\bf p}-{\bf p}_{j,k}),
\end{eqnarray}
where $\Omega_{\Delta p}({\bf p})=1/(\Delta p)^s$ if $p_l\in [p_l-\Delta p/2,
p_l+\Delta p/2]$ for $l=1,\dots ,s$, and is zero otherwise. It follows that
smoothing Eq. ({\ref{unsmo}) over a classically small interval in momentum
gives
\begin{equation}
C_{{\bf n},{\bf m}}^{j,k}({\bf x})\sim\hbar^{s/2}|{\rm det}\frac{\partial^2R_j({\bf
q},{\bf I}_{\bf n})}{\partial {\bf
q}\partial {\bf I}}{\rm det}\frac{\partial^2R_k({\bf q},{\bf I}_{\bf m})}{\partial {\bf
q}\partial {\bf I}}|^{1/2}e^{i(R_j({\bf
q},{\bf I}_{\bf n})-R_k({\bf q},{\bf I}_{\bf m}))/\hbar}
\Omega_{\Delta p}({\bf p}-{\bf p}_{j,k})e^{i(\nu_j-\nu_k)\pi/2}.
\end{equation}

Further smoothing $C_{{\bf n},{\bf m}}^{j,k}({\bf x})$ 
over intervals of length $\Delta q$ about each coordinate variable $q_l$,
$l=1,\dots s$, one can show that the $C_{{\bf n},{\bf m}}^{j,k}({\bf x})$ 
contribution to $\rho_{{\bf n},{\bf m}}^w({\bf x})$ is proportional to 
\begin{equation}
\Pi_{l=1}^s~{\rm sinc}[(p_j^l-p_k^l)\Delta q/2\hbar]
\label{sinc}
\end{equation}
where $p_j^l$ is the $l^{th}$ component of ${\bf p}_j({\bf q},{\bf I})$. 
Thus, for $j\neq k$ Eq. (\ref{sinc}) is $O(h^s)$, while for $j=k$ Eq. (\ref{sinc})
is O(1). Thus, we have shown that smoothing over phase space effectively
eliminates contributions from the $j\ne k$ terms.

Focusing now on the $j=k$ terms, and dropping the subscript $j$, it follows that the diagonal contributions take the form
\begin{equation}
C_{{\bf n},{\bf m}}^{j,j}({\bf x})\sim\hbar^{s/2}|{\rm det}\frac{\partial^2R({\bf
q},{\bf I}_{\bf n})}{\partial {\bf
q}\partial {\bf I}}{\rm det}\frac{\partial^2R({\bf q},{\bf I}_{\bf m})}{\partial {\bf
q}\partial {\bf I}}|^{1/2}e^{i(R({\bf
q},{\bf I}_{\bf n})-R({\bf q},{\bf I}_{\bf m}))/\hbar}\delta({\bf p}-\frac{1}{2}[{\bf p}({\bf q},
{\bf I}_{\bf n})+{\bf p}({\bf q},{\bf I}_{\bf m})]),
\end{equation}
where we have replaced $\Omega_{\Delta p}({\bf p}-\frac{1}{2}[{\bf p}({\bf q},
{\bf I}_{\bf n})+{\bf p}({\bf q},{\bf I}_{\bf m})])$ by the
delta function $\delta({\bf p}-\frac{1}{2}[{\bf p}({\bf q},
{\bf I}_{\bf n})+{\bf p}({\bf q},{\bf I}_{\bf m})])$ 
since we are free to choose $\Delta p$ small.
Assume now that there is an action ${\bf I}_{{\bf n},{\bf m}}$ such that
${\bf p}({\bf q},{\bf I}_{{\bf n},{\bf m}})\sim \frac{1}{2}[{\bf p}({\bf q},
{\bf I}_{\bf n})+{\bf p}({\bf q},{\bf I}_{\bf m})]$ (see Ref.
\cite{jaffe2}), and we also assume that 
${\bf I}_{\bf n}$, ${\bf I}_{\bf m}$, and ${\bf I}_{\bf n,m}$ satisfy relation (\ref{small}), then it readily follows that
\begin{eqnarray}
\frac{R({\bf q},{\bf I}_{\bf
n})-R({\bf q},{\bf I}_{\bf m})}{\hbar}&=&\frac{R({\bf
q},{\bf I}_{{\bf n},{\bf m}}-({\bf I}_{{\bf n},{\bf m}}-{\bf I}_{\bf n}))-R({\bf
q},{\bf I}_{{\bf n},{\bf m}}-({\bf I}_{{\bf n},{\bf m}}-{\bf I}_{\bf
m}))}{\hbar}\nonumber \\
&\sim &\frac{\partial R({\bf q},{\bf I}_{{\bf n},{\bf m}})}{\partial
{\bf I}}\cdot ({\bf n}-{\bf m})
\end{eqnarray}
by Taylor expansion of $R({\bf q},{\bf I}_{\bf n})$ and $R({\bf
q},{\bf I}_{\bf m})$ about ${\bf I}_{{\bf n},{\bf m}}$. 
Further since $\frac{\partial R({\bf q},{\bf I})}{\partial
{\bf I}}=\mbox{\boldmath $\theta$}$, we obtain
\begin{equation}
C_{{\bf n},{\bf m}}^{j,j}({\bf x})\sim \hbar^{s/2}|{\rm det}\frac{\partial^2R({\bf
q},{\bf I}_{\bf n})}{\partial {\bf
q}\partial {\bf I}}{\rm det}\frac{\partial^2R({\bf q},{\bf I}_{\bf m})}{\partial {\bf
q}\partial {\bf I}}|^{1/2}
\delta ({\bf p}-{\bf p}({\bf q},{\bf I}_{{\bf n},{\bf m}}))e^{i({\bf n}-{\bf
m})\cdot \mbox{\boldmath $\theta$}({\bf x})}.
\end{equation}
Finally, summing over the different sheets $j$ [Eq. (\ref{always})], assuming that for small $h$
\begin{equation}
|{\rm det}\frac{\partial^2R({\bf
q},{\bf I}_{\bf n})}{\partial {\bf
q}\partial {\bf I}}{\rm det}\frac{\partial^2R({\bf q},{\bf I}_{\bf m})}{\partial {\bf
q}\partial {\bf I}}|^{1/2}
\sim |{\rm det}\frac{\partial^2R({\bf q},{\bf I}_{{\bf n},{\bf m}})}{\partial {\bf
q}\partial {\bf I}}|,
\end{equation}
and changing variables in the delta function we have
\begin{equation}
\rho_{{\bf n},{\bf m}}^w({\bf x})\sim \hbar^{s/2}\delta 
({\bf I}({\bf x})-{\bf I}_{{\bf n},{\bf m}})e^{i({\bf n}-{\bf
m})\cdot \mbox{\boldmath $\theta$}({\bf x})},
\end{equation}
which is the correct correspondence limit. In particular, 
$\rho_{{\bf n},{\bf m}}^w$ becomes the classical Liouville eigenfunction
[Eq. (\ref{cefs})] with ${\bf I}={\bf I}_{{\bf n},{\bf m}}$ and 
${\bf k}=({\bf n}-{\bf m})$ in the classical limit. Two things are
noteworthy.  
First, an average over $\Delta p \Delta q$ was necessary to eliminate
highly oscillatory terms in the $j\neq k$ contribution to Eq.
(\ref{always}).
Second, under these circumstances the essential singularities 
in the original expression cancel.
  
\section{Classical Liouville Spectral Decomposition}
\label{CLSD}

Attempts to follow a similar correspondence approach for chaotic
systems fails. That is, Eq. (\ref{rhol}) is not true insofar as individual
eigendistributions $\rho^w_{n,m}({\bf x})$ do not have proper
classical limits.  Examination of our motivating equations 
(\ref{qexp}) and (\ref{cexp}) suggests that a one to one correspondence
of quantum to classical eigenfunctions is not necessary to establish
quantum-classical correspondence as $h \rightarrow 0$. Rather, we
require a relation between the quantities in brackets in Eq. 
(\ref{qexp}) and (\ref{cexp}) as $h \rightarrow 0$. In this section
this idea is quantified, for classical systems, by introducing the
classical spectral projectors which we subsequently relate
\cite{chacor}, in the $h \rightarrow 0$ limit, to the analogous
quantum projectors. In addition, we provide a general approach to
computing the classical projectors for arbitrary systems.

The classical projectors will be seen to offer a better method of
describing chaotic systems than do the classical Liouville
eigenfunctions introduced above [Eq. (\ref{grail})]. Specifically,
they are not based on the arbitrary set $\chi_\ell$, they are
completely defined as opposed to the $\rho_{E, \lambda}^\ell({\bf x})$
which are defined to within an overall phase, and they offer a means of
separating the spectrum into continuous and discrete components\cite{spec}.

We also extend the approach and separate the singular
spectrum of the periodic orbits from the rest of the continuous
spectrum. The resulting periodic orbit spectral projection operators
are used in the companion paper on correspondence in chaotic
systems \cite{chacor}.

To develop an energy/frequency spectral decomposition we must isolate
the fixed frequency and energy components of
an arbitrary time evolving phase space probability density $\rho ({\bf
x},t)$ for a conservative classical Hamiltonian system of $s$ degrees
of freedom. Phase space probability densities $\rho({\bf x},t)=\rho({\bf X}({\bf
x},-t),0)$ evolve in classical
mechanics in accord with the
classical Liouville equation, Eq. (\ref{what}).
Consider that
\begin{equation}
\rho({\bf x},t)=\int
d{\bf x}_0 ~\rho({\bf x}_0,0)~\delta({\bf x}_0-{\bf X}({\bf x},-t))
\end{equation}
where $\delta({\bf x}_0-{\bf X}({\bf x},-t))$ is the kernel of the
Liouville propagator,
which can be rewritten as
\begin{equation}
\rho({\bf x},t)=\int_{-\infty}^{\infty} dt'\delta(t-t')\int
d{\bf x}_0~\rho({\bf x}_0,0)~\delta({\bf x}_0-{\bf X}({\bf x},-t')).
\end{equation}
Making the replacement $\delta(t-t')=\frac{1}{2\pi}\int_{-\infty}^{\infty} d\lambda~
e^{-i\lambda(t-t')}$ we obtain
\begin{equation}
\rho({\bf x},t)=\int_{-\infty}^{\infty} dt'\frac{1}{2\pi}\int_{-\infty}^{\infty} d\lambda
e^{-i\lambda(t-t')}\int
d{\bf x}_0 ~\rho({\bf x}_0,0)~\delta({\bf x}_0-{\bf X}({\bf x},-t')).
\end{equation}
Finally inserting a closure relation for the energy 
$1=\int_0^{\infty} dE~\delta(E-H({\bf x}))$ gives
\begin{equation}
\rho({\bf x},t)=\int_0^{\infty} dE\int_{-\infty}^{\infty} d\lambda ~e^{-i\lambda t}\int
d{\bf x}_0~\rho({\bf x}_0,0)~\Upsilon_{E,\lambda}({\bf x};{\bf x}_0),
\label{fexp}
\end{equation}
where
\begin{equation}
\Upsilon_{E,\lambda}({\bf x};{\bf x}_0)=\frac{1}{2\pi}\delta(E-H({\bf
x}_0))\int_{-\infty}^{\infty}dt^{\prime}e^{i\lambda
t^{\prime}}\delta({\bf x}_0-{\bf X}({\bf x},-t^{\prime})).
\label{d2}
\end{equation}
Equation (\ref{fexp}) provides an expansion of $\rho({\bf x},t)$ in terms
of the classical spectral projection operators $\Upsilon_{E,\lambda}$
which are expressed, in Eq. (\ref{d2}), in terms of the dynamics.
Specifically,
$\Upsilon_{E,\lambda}({\bf x};{\bf x}_0)$ is the Fourier transform of
the kernel of the Liouville propagator, restricted to the energy shell.  
Equation (\ref{fexp}) is
a restatement of Eq. (\ref{cexp}) which emphasizes the classical
spectral projection operators as being central to the analysis of the
dynamical evolution of distributions $\rho({\bf x},t)$.

The distribution $\Upsilon_{E,\lambda}({\bf x};{\bf x}_0)$, as a
function of ${\bf x}$, is
an eigenfunction of the Liouville operator with eigenvalue $\lambda$. 
 To see this consider that
propagation with respect to the ${\bf x}$ variables gives
\begin{eqnarray}
e^{-iL_c t}\Upsilon_{E,\lambda}({\bf x};{\bf x}_0)&=&\frac{1}{2\pi}\delta(E-H({\bf
x}_0))\int_{-\infty}^{\infty}dt^{\prime}e^{i\lambda
t^{\prime}}\delta({\bf x}_0-{\bf X}({\bf x},-t^{\prime}-t))\nonumber
 \\
&=&e^{-i\lambda t}\Upsilon_{E,\lambda}({\bf x};{\bf x}_0),
\label{peig}
\end{eqnarray}
where we have made the change of variable $t'+t\rightarrow t'$. 
Similarly, with $L_c({\bf x}_0)=i\frac{\partial H({\bf x}_0)}{\partial
{\bf x}_0}\cdot J\cdot \widetilde{\frac{\partial}{\partial {\bf x}_0}}$ 
operating on the ${\bf x}_0$ variables,
$L_c\Upsilon_{E,\lambda}({\bf x};{\bf
x}_0)=-\lambda\Upsilon_{E,\lambda}({\bf x};{\bf x}_0)$. [Here $J=\left( \begin{array}{cc}
0&-I \\
I &0
\end{array} \right)$ is the $2s\times 2s$ dimensional symplectic
matrix\cite{gold}]. Thus, $\Upsilon_{E,\lambda}({\bf x};{\bf
x}_0)$ is an eigenfunction of $L_c$ with eigenvalue $\lambda$
($-\lambda$) in the
${\bf x}$ (${\bf x}_0$) variables. These properties are consequences
of the fact that 
$\Upsilon_{E,\lambda}({\bf x};{\bf x}_0)=\Upsilon_{E,-\lambda}({\bf
x}_0;{\bf x})$,
which is in turn a consequence of the fact that $L_c$ is unitarily
equivalent to $-L_c$.

The decomposition in Eq. (\ref{fexp}) holds for all distributions
and all phase space points ${\bf x}$,
including periodic orbits. To show this,
suppose that ${\bf x}$ lies on a periodic orbit.  Then by
Eq. (\ref{d2})
\begin{eqnarray}
\Upsilon_{E,\lambda}({\bf x};{\bf x}_0)&=&\frac{1}{2\pi}\delta(E-H({\bf x}_0))
\int_{-\infty}^{\infty}dt^{\prime}e^{i\lambda
t^{\prime}}\delta({\bf x}_0-{\bf X}({\bf x},-t^{\prime}))\nonumber \\
&=&\frac{1}{2\pi}\delta(E-H({\bf x}_0))
\sum_{j=-\infty}^{\infty}e^{ij\lambda\tau}\int_{-\tau/2}^{\tau/2}dt^{\prime}e^{i\lambda
t^{\prime}}\delta({\bf x}_0-{\bf X}({\bf x},-t^{\prime}))\nonumber \\
&=&\delta(E-H({\bf x}_0))
\sum_{j=-\infty}^{\infty}\delta(\lambda-2\pi j/\tau)\frac{1}{\tau}\int_{-\tau/2}^{\tau/2}dt^{\prime}e^{2\pi
ij
t^{\prime}/\tau}\delta({\bf x}_0-{\bf X}({\bf x},-t^{\prime}))
\end{eqnarray}
where $\tau$ is the fundamental period of the orbit. Thus we have
\begin{equation}
\Upsilon_{E,\lambda}({\bf x};{\bf x}_0)=\delta(E-H({\bf x}_0))\sum_{j=-\infty}^{\infty}\delta(\lambda-2\pi j/\tau)\frac{1}{\tau}\int_{-\tau/2}^{\tau/2}dt^{\prime}e^{2\pi
ij
t^{\prime}/\tau}\delta({\bf x}_0-{\bf X}({\bf x},-t^{\prime}))
\label{uppo}
\end{equation}
on periodic orbits.
Now inserting this
expression into the right hand side of Eq. (\ref{fexp}) and performing
the integrals over $\lambda$ and $E$ gives
\begin{eqnarray}
\rho({\bf x},t)=&&\int
d{\bf x}_0 ~\rho({\bf x}_0,0)~\sum_{j=-\infty}^{\infty}\frac{1}{\tau}\int_{-\tau/2}^{\tau/2}dt^{\prime}e^{2\pi
ij
(t^{\prime}-t)/\tau}\delta({\bf x}_0-{\bf X}({\bf x},-t^{\prime}))\nonumber \\
&&=\int
d{\bf x}_0 ~\rho({\bf x}_0,0)~\sum_{j=-\infty}^{\infty}\int_{-\tau/2}^{\tau/2}dt^{\prime}\delta(t^{\prime}-t-j\tau)\delta({\bf x}_0-{\bf X}({\bf x},-t^{\prime}))\nonumber \\
&&=\int
d{\bf x}_0 ~\rho({\bf x}_0,0)~\delta({\bf x}_0-{\bf X}({\bf x},-t)).
\end{eqnarray}
Performing the integrals over $d{\bf x}_0$ gives the
probability density on the point ${\bf x}$ of the periodic
orbit, as required. Thus, the expansion (\ref{fexp}) and its associated
closure relation hold for points on periodic orbits.

  Since Eq. (\ref{fexp}) holds for all $\rho$ and all ${\bf
  x}$ it implies the existence
of the closure relation
\begin{equation}
\int_0^{\infty} dE\int_{-\infty}^{\infty} d\lambda ~\Upsilon_{E,\lambda}({\bf x};{\bf
x}_0) =\delta({\bf x}_0-{\bf x}).
\end{equation}
 Two unsatisfactory issues require more careful
attention:  First, there is no clear separation of
the point spectrum from the continuous spectrum\cite{spec}. 
Second, a comparison of Eq. (\ref{fexp}) with Eq. (\ref{cexp}) suggests
 that $\Upsilon_{E,\lambda}({\bf x}, {\bf x}_0)$ should relate 
to the Liouville eigenfunctions $\rho_{E,\lambda}^\ell$ as a product of an
eigenfunction in ${\bf x}_0$ times
an eigenfunction in ${\bf x}$. These issues are addressed below where
we treat the integrable and chaotic cases separately.

\subsection{Integrable Systems}
\label{IS2}

To demonstrate the computation of $\Upsilon$ consider integrable
systems.  Here we change to action angle variables ${\bf x}=({\bf
I},\mbox{\boldmath $\theta$})$ and ${\bf x}_0=({\bf
I}_0,\mbox{\boldmath $\theta$}_0)$ so that $\Upsilon_{E,\lambda}$
assumes the form
\begin{eqnarray}
\Upsilon_{E,\lambda}({\bf x};{\bf x}_0)&=&\frac{1}{2\pi}\delta(E-H({\bf I}_0))
\int_{-\infty}^{\infty}dt^{\prime}e^{i\lambda
t^{\prime}}\delta({\bf I}_0-{\bf I})\delta((\mbox{\boldmath
$\theta$}_0-\mbox{\boldmath $\theta$} +\mbox{\boldmath $\omega$}({\bf I})
t^{\prime})~{\rm mod}~2\pi)\nonumber \\
&=&\frac{1}{2\pi}\delta(E-H({\bf I}_0))
\int_{-\infty}^{\infty}dt^{\prime}e^{i\lambda
t^{\prime}}\delta({\bf I}_0-{\bf I})
\frac{1}{(2\pi)^s}\sum_{{\bf k}\in {\bf Z}^s}e^{-i{\bf k}\cdot[\mbox{\boldmath $\theta$}_0-\mbox{\boldmath $\theta$} +\mbox{\boldmath $\omega$}({\bf I})
t^{\prime}]},
\label{wherever}
\end{eqnarray}
where we have used the identity\cite{vlad}
\begin{equation}
\sum_{{\bf j}\in {\bf Z}^s}\delta(\mbox{\boldmath $\theta$}_0-\mbox{\boldmath $\theta$} +\mbox{\boldmath $\omega$}({\bf I})
t^{\prime}-2\pi {\bf j})=\frac{1}{(2\pi)^s}\sum_{{\bf k}\in {\bf Z}^s}e^{-i{\bf k}\cdot[\mbox{\boldmath $\theta$}_0-\mbox{\boldmath $\theta$} +\mbox{\boldmath $\omega$}({\bf I})
t^{\prime}]}.
\label{idvlad}
\end{equation}
In terms of the Liouville eigendistributions $\rho_{{\bf I}',{\bf n}}({\bf
I},\mbox{\boldmath $\theta$})$ [Eq. (\ref{cefs})]
Eq. (\ref{wherever}) becomes
\begin{equation}
\Upsilon_{E,\lambda}({\bf x};{\bf x}_0)=\frac{1}{2\pi}\delta(E-H({\bf I}_0))
\sum_{{\bf n}}\int d{\bf I}'\delta(\lambda-{\bf n}\cdot\mbox{\boldmath $\omega$}({\bf I}'))\rho^*_{{\bf I}',{\bf n}}({\bf
I}_0,\mbox{\boldmath $\theta$}_0)\rho_{{\bf I}',{\bf n}}({\bf
I},\mbox{\boldmath $\theta$}).
\label{eq 44}
\end{equation}
Inserting Eq. (\ref{eq 44}) into Eq. (\ref{fexp}) gives the expansion
\begin{eqnarray}
&&\rho({\bf x},t)=\sum_{{\bf n}}\int d{\bf I}'e^{-i{\bf
n}\cdot\mbox{\boldmath $\omega$}({\bf I}')t}\int d{\bf I}_0d 
\mbox{\boldmath $\theta$}_0~\rho({\bf I}_0,
\mbox{\boldmath $\theta$}_0,0) \left[ \rho^*_{{\bf I}',{\bf n}}({\bf I}_0,
\mbox{\boldmath $\theta$}_0)\rho_{{\bf I}',{\bf n}}
({\bf I},\mbox{\boldmath $\theta$})\right].
\end{eqnarray}
This result suggests a natural definition of spectral projection operators for integrable
systems, i.e.,
\begin{equation}
{\cal Y}_{{\bf I}',{\bf n}}({\bf x};
{\bf x}_0)= \rho^*_{{\bf I}',{\bf n}}
({\bf x}_0)\rho_{{\bf I}',{\bf n}}({\bf x}),
\label{sprod}
\end{equation}
which takes the
form of a product of Liouville eigenfunctions. These spectral projection
operators satisfy the following symmetry, closure and orthonormality
relations:
\begin{equation}
{\cal Y}_{{\bf
I}',{\bf n}}({\bf 
x};{\bf 
x}_0)={\cal Y}_{{\bf
I}',-{\bf n}}({\bf 
x}_0;{\bf 
x}),
\end{equation}
\begin{equation}
\int d{\bf I}'~{\cal Y}_{{\bf I}',0}({\bf x};{\bf x}_0)+ \sum_{{\bf
n}\neq 0} \int d{\bf I}'~{\cal Y}_{{\bf I}',{\bf n}}({\bf x};{\bf x}_0) 
=\delta ({\bf x}-{\bf x}_0),
\label{rclo}
\end{equation}
and
\begin{equation}
\int d{\bf x}~{\cal Y}_{{\bf I}'',{\bf n}'}^*({\bf x};{\bf
x}_0^{\prime})~{\cal Y}_{{\bf I}',{\bf n}}({\bf x};{\bf x}_0)=
\delta_{{\bf n},{\bf n}'}\delta({\bf I}''-{\bf I}'){\cal Y}_{{\bf I}',{\bf n}}({\bf x}_0^{\prime};{\bf
x}_0).
\end{equation}
Further, the  decomposition in Eq. (\ref{rclo})
clearly displays the separation of point and continuous spectrum.

\subsection{Chaotic Systems}
\label{CS2}

The time evolution of a phase space distribution for a
chaotic system which is hyperbolic can be decomposed in the following
fashion (see
Appendix B)
\begin{equation}
\rho({\bf x},t)=\int_0^{\infty} dE\int
d{\bf x}_0~\rho({\bf x}_0,0)~\Upsilon_E({\bf x};{\bf x}_0)
+\int_0^{\infty} dE\int\!\!\!\!\!\!- d\lambda ~e^{-i\lambda t}\int
d{\bf x}_0 ~\rho({\bf x}_0,0)~\Upsilon_{E,\lambda}({\bf x};{\bf x}_0),
\label{cdcomp}
\end{equation}
where $\Upsilon_{E,\lambda}({\bf x};{\bf x}_0)$ is given by Eq.
(\ref{d2}) and
\begin{equation}
\Upsilon_E({\bf x};{\bf x}_0)=
\frac{\delta(E-H({\bf x}_0))\delta(E-H({\bf x}))}{\int d{\bf x}'\delta
(E-H({\bf x}'))}.
\label{d1}
\end{equation}
Here $\int\!\!\!\!\!- d\lambda$ denotes the integral over $\lambda$ with the
point spectrum at $\lambda=0$ removed, i.e., $\int\!\!\!\!\!- d\lambda =\int_{-\infty}^{\infty}
d\lambda P_E$, where $P_E\rho=[\rho-\langle \rho\rangle_E]$, and
\begin{equation}
\langle
\rho\rangle_E=\frac{\int d{\bf x}~\delta(E-H({\bf x}))~\rho({\bf
x},0)}{\int d{\bf x}'\delta
(E-H({\bf x}'))}
\label{mcav}
\end{equation}
is the microcanonical average of
$\rho$. The operator $P_E$ projects distributions onto the continuous
spectrum part of the Hilbert space. Properties such as
the system is ergodic ($\lambda=0$ is a nondegenerate point
eigenvalue with a corresponding eigenfunction which is uniform over the energy
surface), weak mixing (no point eigenvalues other than $\lambda=0$),
and positive
Kolmogorov entropy (i.e., the remainder of the spectrum
is continuous) have been incorporated in Eq. (\ref{cdcomp})
in an obvious way. Since the component of $\rho({\bf x})$ with energy $E$
is given by
\begin{equation}
\int d{\bf x}_0 ~\rho({\bf x}_0,0)\Upsilon_{E}({\bf
x};{\bf x}_0)=\langle
\rho\rangle_E\delta(E-H({\bf x}))
\end{equation}
we see that $\Upsilon_{E}$ projects $\rho({\bf x})$
onto the point spectrum part of the Hilbert space. Similarly,
$\Upsilon_{E,\lambda}$ projects an arbitrary
distribution onto the continuous spectrum part of the Hilbert space
associated with a frequency $\lambda$ and energy $E$. 
We further note that this set of projection operators is complete
\begin{equation}
\int_0^{\infty} dE~\Upsilon_{E}({\bf x};{\bf x}_0)+\int_0^{\infty} dE\int\!\!\!\!\!\!- d\lambda ~\Upsilon_{E,\lambda}({\bf x};{\bf x}_0)
=\delta({\bf x}-{\bf x}_0),
\label{rel6}
\end{equation}
and orthogonal:
\begin{equation}
\int d{\bf x}~\Upsilon_{E^{\prime}}^{*}({\bf x};{\bf x}_0^{\prime})~\Upsilon_{E}({\bf x};{\bf x}_0)=
\delta(E-E^{\prime})\Upsilon_{E}({\bf x}_0^{\prime};{\bf x}_0),
\label{rel7}
\end{equation}
and
\begin{equation}
\int d{\bf x}~\Upsilon_{E^{\prime},\lambda^{\prime}}^*({\bf x};{\bf x}_0^{\prime})~\Upsilon_{E,\lambda}({\bf x};{\bf x}_0)=
\delta(E-E^{\prime})\delta(\lambda-\lambda^{\prime})\Upsilon_{E,\lambda}({\bf x}_0^{\prime};{\bf x}_0).
\label{rel8}
\end{equation}
[These relations are proven in Appendix B.] The distributions
$\Upsilon_{E}$ and $\Upsilon_{E,\lambda}$ are thus the kernels of orthogonal
stationary and nonstationary spectral
projection operators for the classical Liouville spectrum. We have
thus achieved, in this approach, the separation of point and
continuous spectrum.  It remains to examine the relationship of $\Upsilon_{E,\lambda}$ 
to the Liouville eigenfunctions. 

To do so insert Eq. (\ref{le1}) into the definition of 
$\Upsilon_{E,\lambda}$ [Eq.(\ref{d2})] to obtain
\begin{equation}
\Upsilon_{E,\lambda}({\bf x};{\bf x}_0)=\frac{1}{2\pi}\delta(E-H({\bf x}_0))\delta(E-H({\bf x}))e^{-i\lambda(\tau({\bf
x}_0)-\tau({\bf
x}))}\delta(\mbox{\boldmath $\eta$}({\bf
x}_0)-\mbox{\boldmath $\eta$}({\bf
x})).
\label{eq108}
\end{equation}
Inserting Eq. (\ref{le2}) in Eq. (\ref{eq108}) and rewriting in terms of the definition of the chaotic Liouville
eigenfunctions [Eq. (\ref{grail})]
gives
\begin{equation}
\Upsilon_{E,\lambda}({\bf x};{\bf x}_0)=\sum_\ell\rho_{E,\lambda}^{\ell*}
({\bf x}_0)\rho_{E,\lambda}^\ell ({\bf x}).
\label{decompos}
\end{equation}
We also define stationary Liouville eigenfunctions, $\rho_E({\bf
x})$, via
\begin{equation}
\rho_E({\bf x})=\frac{\delta(E-H({\bf x}))}{[\int d{\bf x}'\delta
(E-H({\bf x}'))]^{1/2}},
\label{rhoe}
\end{equation}
such that
\begin{equation}
\Upsilon_E({\bf x};{\bf x}_0)=\rho_E^*({\bf
x}_0)\rho_E({\bf x}).
\label{upee}
\end{equation}
The distributions $\rho_E({\bf x})$ and $\rho_{E,\lambda}^{\ell}({\bf x})$
are zero off the energy shell $E=H({\bf x})$. The eigendistributions
$\rho_E({\bf x})$ are stationary and uniform over the energy shell and
belong to the point spectrum.
Distributions $\rho_{E,\lambda}^{\ell}({\bf x})$ are nonuniform and
stationary for $\lambda=0$, and nonuniform and nonstationary for
$\lambda\neq 0$. The distributions $\rho_{E,\lambda}^{\ell}({\bf x})$
belong to the continuous spectrum. Finally, note that
$\rho_{E}({\bf x})$ and $\rho_{E,\lambda}^{\ell}({\bf x})$ are supported
over the entire ($2s-1$)-dimensional energy surface $E=H({\bf x})$, in
contrast to the distributions
$\rho_{{\bf I}',{\bf k}}({\bf x})$ for integrable systems which are
supported only on the $s$-dimensional torus ${\bf I}'={\bf I}({\bf x})$.
In other words, every orbit of the energy surface $E=H({\bf x})$
contributes to the construction of a chaotic eigenfunction.

Comparing
\begin{equation}
\Upsilon_{E}({\bf x};{\bf x}_0)=\rho_E^*({\bf x}_0)\rho_E({\bf x})
\end{equation}
with
\begin{equation}
\Upsilon_{E,\lambda}({\bf x};{\bf
x}_0)=\sum_{\ell}\rho_{E,\lambda}^{\ell *}({\bf
x}_0)\rho_{E,\lambda}^{\ell} ({\bf x})
\end{equation}
we see that the stationary spectral projection operators
$\Upsilon_{E}$ are simple products of Liouville eigenfunctions while
$\Upsilon_{E,\lambda}$ are composed of a sum of products of Liouville
eigenfunctions. This structure, i.e., a sum over products, has
important implications for the
correspondence of the nonstationary chaotic quantum Liouville eigenfunctions
\cite{chacor}. Hence we shall
prove here that $\Upsilon_{E,\lambda}({\bf x};{\bf x}_0)$
cannot be rewritten as a simple product of some new, as yet undiscovered,
eigenfunctions. 

Suppose that we can write
\begin{equation}
\Upsilon_{E,\lambda}({\bf x};{\bf x}_0)=\Omega_{E,\lambda}^*({\bf
x}_0)\Omega_{E,\lambda}({\bf x})
\label{prod1}
\end{equation}
for some unknown distributions $\Omega_{E,\lambda}({\bf x})$.
Then, in order to satisfy the orthogonality relation [Eq.
(\ref{rel8})] we must have
\begin{equation}
\int
d{\bf x}~\Omega_{E^{\prime},\lambda^{\prime}}^*({\bf
x})\Omega_{E,\lambda}({\bf x})=
\delta(E-E^{\prime})\delta(\lambda-\lambda^{\prime}).
\label{prod2}
\end{equation}
Equations (\ref{prod1}) and (\ref{prod2}) along with the definition of
$\Upsilon_{E,\lambda}$ then imply that
\begin{eqnarray}
\delta(E-E^{\prime})\delta(\lambda-\lambda^{\prime})\Omega_{E,\lambda}({\bf
x})&=&\frac{1}{2\pi}\delta(E-H({\bf x}))\int_{-\infty}^{\infty}dt^{\prime}e^{i\lambda
t^{\prime}}\Omega_{E^{\prime},\lambda^{\prime}}({\bf X}({\bf x},-t^{\prime}))\nonumber
\\
&=&\delta(E-H({\bf x}))\delta(\lambda-\lambda^{\prime})\Omega_{E^{\prime},\lambda^{\prime}}({\bf x}).
\end{eqnarray}
Integrating over $\lambda^{\prime}$ and $E^{\prime}$ gives
\begin{equation}
\Omega_{E,\lambda}({\bf x})
=\delta(E-H({\bf x}))\int dE^{\prime}~\Omega_{E^{\prime},\lambda}({\bf x}),
\end{equation}
which implies that
$\Omega_{E,\lambda}({\bf x})\propto\delta(E-H({\bf x}))$. 
Note that $\Upsilon_{E,\lambda}({\bf x};{\bf x}_0)=0$ unless
${\bf x}_0$ and ${\bf x}$ lie on the same
trajectory. Thus, if
$\Omega_{E,\lambda}({\bf x})=\delta(E-H({\bf x}))N_{\lambda}({\bf x})$
then
$N_{\lambda}^*({\bf x}_0)N_{\lambda}({\bf x})=0$
unless ${\bf x}_0$ and ${\bf x}$ lie on the same
trajectory. However, if ${\bf x}_0$ and
${\bf x}$ lie on the same trajectory then
$H({\bf x})=H({\bf x}_0)$ and so
\begin{eqnarray}
\Omega_{E,\lambda}^*({\bf x}_0)\Omega_{E,\lambda}({\bf x})&=&\delta(E-H({\bf x}_0))N_{\lambda}^*({\bf x}_0)\delta(E-H({\bf x}))N_{\lambda}({\bf x})
\nonumber \\
&=&\delta(E-H({\bf x}))^2N_{\lambda}^*({\bf x}_0)N_{\lambda}({\bf x})
\end{eqnarray}
so that
$\Omega_{E,\lambda}^*({\bf x}_0)\Omega_{E,\lambda}({\bf x})$
would not even be integrable in a delta function sense. Thus
$\Upsilon_{E,\lambda}$ cannot be written as a simple product of
Liouville eigenfunctions.

\subsubsection{Periodic Orbits}

Periodic orbits play an important role in modern theories of
quantization in chaotic systems.
Given that Eqs. (\ref{fexp}) and (\ref{cdcomp}) hold for points on
periodic orbits, and given that hyperbolic systems have a countable
number of isolated periodic orbits for any given energy, it is
also possible to separate the singular spectrum associated
with the periodic orbits from the rest of the continuous spectrum. To do
this we define periodic orbit distributions (extracted from Eq. (\ref{uppo})):
\begin{equation}
\Upsilon_{E,j}^k({\bf x};{\bf x}_0)=\delta(E-H({\bf x}_0))
\frac{1}{\tau_k}\int_{-\tau_k/2}^{\tau_k/2}dt^{\prime}e^{2\pi
ij
t^{\prime}/\tau_k}\delta({\bf x}^k-{\bf X}({\bf x},-t^{\prime}))
\label{poeq}
\end{equation}
where ${\bf x}^k$ is a point on the $k^{{\rm th}}$
periodic orbit at energy $E$. [Here again $\tau_k$ is the fundamental period
of orbit $k$]. Note that
the distributions $\Upsilon_{E,0}^k$ have the property that
\begin{equation}
\frac{\int d{\bf x}d{\bf x}_0~\Upsilon_{E,0}^k({\bf
x};{\bf x}_0) F({\bf x}_0)G({\bf x})}{\int d{\bf x}'\delta
(E-H({\bf x}'))}=\langle F\rangle_E\langle G\rangle_{E,0}^k
\end{equation}
where $F$ and $G$ are any two observables, $\langle F\rangle_E$
denotes a microcanonical average at energy $E$, and $\langle
G\rangle_{E,j}^k\equiv\frac{1}{\tau_k}\int_{-\tau_k/2}^{\tau_k/2}dt'e^{2\pi
ij
t^{\prime}/\tau_k}~G({\bf
X}({\bf x}^k,t'))$ is the $j^{th}$ Fourier component of $G$ on periodic orbit $k$ at energy $E$.
The total closure relation now takes the form
\begin{eqnarray}
&&\int dE~\Upsilon_{E}({\bf x};{\bf x}_0)+\int dE~\sum_j'\sum_{\mbox{\boldmath $\eta$}_k\in{\cal P}}\delta_{\mbox{\boldmath $\eta$}({\bf x}_0),\mbox{\boldmath $\eta$}_k}\Upsilon_{E,j}^k({\bf x};{\bf
x}_0)+\int dE\int\!\!\!\!\!\!- d\lambda ~\Upsilon_{E,\lambda}({\bf
x};{\bf x}_0)\nonumber \\
&&=\delta({\bf x}-{\bf x}_0),
\label{tdecomp}
\end{eqnarray}
where ${\cal P}\equiv\{\mbox{\boldmath
$\eta$}|\mbox{\boldmath $\eta$} ~{\rm labels~a~periodic~orbit}\}$
and the $\mbox{\boldmath $\eta$}$ variables are those which we introduced earlier, and where $\int\!\!\!\!\!- d\lambda$ now denotes the integral over $\lambda$ with the
point and singular spectrum removed, i.e., $\int\!\!\!\!\!- d\lambda=\int
d\lambda ~S_EP_E$, where
\begin{equation}
S_E\rho({\bf x})=\rho({\bf x})-\sum_j\sum_{\mbox{\boldmath $\eta$}_k\in{\cal P}}\delta_{\mbox{\boldmath $\eta$}({\bf x}),\mbox{\boldmath $\eta$}_k}\langle
\rho\rangle_{E,j}^k
\end{equation}
and so 
\begin{equation}
S_EP_E\rho({\bf x})=\rho({\bf
x})-\langle\rho\rangle_E -\sum_j\sum_{\mbox{\boldmath $\eta$}_k\in{\cal P}}\delta_{\mbox{\boldmath $\eta$}({\bf x}),\mbox{\boldmath $\eta$}_k}\langle
[\rho-\langle\rho\rangle_E]\rangle_{E,j}^k. 
\end{equation}
The distributions $\Upsilon_{E,j}^k({\bf x};{\bf
x}_0)$ are asymmetric in the ${\bf x}$, ${\bf x}_0$ variables, and so
a Kronecker delta $\delta_{\mbox{\boldmath $\eta$}({\bf
x}_0),\mbox{\boldmath $\eta$}_k}$ is introduced as a factor in the
singular spectrum term of  Eq. (\ref{tdecomp}) in order to restore
the proper symmetry with respect to ${\bf x}$ and ${\bf x}_0$. The
prime over the sum on $j$
denotes that the point spectrum has been removed, i.e., $\sum_j'=\sum_j P_E$.

One last aspect of definition (\ref{poeq}) should be emphasized. For
$j=0$ we may write
\begin{eqnarray}
\Upsilon_{E,0}^k({\bf x};{\bf x}_0)&=&\delta(E-H({\bf x}_0))\frac{1}{\tau_k}\int_{-\tau_k/2}^{\tau_k/2}dt^{\prime}~\delta(\tau({\bf p}^k,{\bf
q}^k)-\tau({\bf p},{\bf
q})+t^{\prime})\cdot\nonumber \\
&&\delta(E-H({\bf x}))\delta(\mbox{\boldmath $\eta$}_k-\mbox{\boldmath $\eta$}({\bf x}))\nonumber \\
&=&\tau_k^{-1}\delta(E-H({\bf x}_0))\delta(E-H({\bf x}))\delta(\mbox{\boldmath $\eta$}_k-\mbox{\boldmath $\eta$}({\bf x})).
\end{eqnarray}
But local coordinates $\mbox{\boldmath $\xi$}_k({\bf
x})$ of the Poincare surface of section\cite{berry3,hannay}, transverse to
periodic orbit $k$, may be introduced such that
\begin{equation}
\delta(\mbox{\boldmath $\eta$}_k-\mbox{\boldmath $\eta$}({\bf x}))=\delta(\mbox{\boldmath $\xi$}_k({\bf x})). 
\label{globloc}
\end{equation}
Note that $\mbox{\boldmath $\eta$}_k-
\mbox{\boldmath $\eta$}({\bf x})\neq \mbox{\boldmath $\xi$}_k({\bf x})$ 
since $\mbox{\boldmath $\eta$}({\bf x})$ are global $\tau$ independent
variables, while $\mbox{\boldmath $\xi$}_k({\bf x})$ are $\tau$ dependent
local variables, i.e., $\mbox{\boldmath $\xi$}_k(H,\tau+\tau_k,\mbox{\boldmath $\eta$})= M_k\mbox{\boldmath $\xi$}_k(H,\tau,\mbox{\boldmath $\eta$})$ where
$M_k$ is the stability matrix of orbit $k$. However, since
\begin{equation}
\delta(\mbox{\boldmath $\xi$}_k(H,\tau+\tau_k,\mbox{\boldmath $\eta$}))= \frac{1}{|M_k|}\delta(\mbox{\boldmath $\xi$}_k(H,\tau,\mbox{\boldmath $\eta$}))
\end{equation}
and $|{\rm det}M_k|=1$, it follows that $\delta(\mbox{\boldmath $\xi$}_k(H,\tau,\mbox{\boldmath $\eta$}))$ is time independent.
Thus, the stationary periodic orbit spectral projection operators can be 
written in the form
\begin{equation}
\Upsilon_{E,0}^k({\bf x};{\bf x}_0)=\tau_k^{-1}\delta(E-H({\bf
x}_0))\delta(E-H({\bf x}))\delta(\mbox{\boldmath $\xi$}_k({\bf x})).
\label{podist}
\end{equation}
This particular form will be of use in the companion paper
\cite{chacor} on correspondence in chaotic systems in which scar
contributions are related to these spectral projectors. Indeed, 
the analysis of the classical dynamics of distributions, contained in this
section,  will prove
central to the analysis of classical-quantum correspondence in chaotic
systems, discussed in the following paper \cite{chacor}.

\section{Summary}
\label{summary}
In summary, we have constructed a coherent framework
for the study of quantum-classical correspondence. We have emphasized
the importance of considering, in each mechanics, the entire set of $2r$
eigenequations in the Liouville picture. We have then shown how this
approach, when combined with previous demonstrations of correspondence
for the Liouville eigenfunctions, allows a complete understanding of
correspondence for integrable systems. General methods were then discussed
for the construction of Liouville spectral decompositions necessary for the
study of correspondence in chaotic systems.
These methods were then employed for the construction of 
Liouville eigenfunctions for chaotic systems. It was also shown that
application of these methods to integrable
systems yields the usual spectral decomposition. Finally, we reviewed the
arguments for the correspondence of Liouville eigenfunctions for
quantum systems with integrable classical analogs and showed that the
primary mechanism of correspondence is the elimination of essential
singularities when there is an averaging over a small range $\Delta p
\Delta q$.

In the following paper \cite{chacor} we show that the correspondence
problem for systems with chaotic classical analog can be treated by
the methods introduced in this paper.

\vspace{1 in}

{\bf Acknowledgements:} We thank the Natural Sciences and Engineering
Research Council of Canada for support of this work.

\section*{Appendix A}
\label{3A}

Here we will prove the orthogonality and completeness relations
(\ref{sdec}) for the chaotic Liouville eigenfunctions
$\rho_{E,\lambda}^{\ell}$. To prove (\ref{sdec}) consider that
\begin{eqnarray}
\int d{\bf x}~\rho_{E',\lambda '}^{\ell'*}({\bf x}_0)\rho_{E,\lambda}^{\ell}
({\bf x}) &=&\int dHd\tau d\mbox{\boldmath
$\eta$}~\frac{1}{2\pi}\int_{-\infty}^{\infty}d\tau
e^{i(\lambda-\lambda ')\tau}\cdot\nonumber \\
&&~\delta(E-E')\delta(H-E)\chi_{{\ell}'}^*(\mbox{\boldmath $\eta$})\chi_{\ell}(\mbox{\boldmath $\eta$})\nonumber \\
&=&\delta_{\ell',\ell}\delta(E'-E)\delta(\lambda '-\lambda)
\end{eqnarray}
where we have used Eq. (\ref{le3}) and the identity
$\delta(x)=\frac{1}{2\pi}\int_{-\infty}^{\infty} dy~e^{ixy}$. 

To show completeness note that
\begin{eqnarray}
&&\sum_{\ell}\int_0^{\infty}dE\int_{-\infty}^{\infty}
d\lambda~\rho_{E,\lambda}^{\ell *} ({\bf x}')\rho_{E,\lambda}^{\ell} ({\bf
x})=\frac{1}{2\pi}\sum_{\ell}\int_0^{\infty}dE\int_{-\infty}^{\infty}
d\lambda~\delta(E-H({\bf x}')\cdot\nonumber \\
&&\delta(E-H({\bf x})e^{i\lambda (\tau({\bf
x})-\tau({\bf x}'))}\chi_{\ell}^*(\mbox{\boldmath $\eta$}({\bf
x}'))\chi_{\ell}(\mbox{\boldmath $\eta$}({\bf
x}))\nonumber \\
&=&\delta(H({\bf x})-H({\bf x}'))\delta(\tau({\bf x})-\tau({\bf x}'))\delta(\mbox{\boldmath $\eta$}({\bf
x})-\mbox{\boldmath $\eta$}({\bf
x}'))\nonumber \\
&=&\delta ({\bf x}-{\bf x}')
\end{eqnarray}
where we used Eq. (\ref{le2}) and the identity
$\delta(x)=\frac{1}{2\pi}\int_{-\infty}^{\infty} dy~e^{ixy}$ to go
from the first line to the second,
and Eq. (\ref{le1}) to go from the second line to the third.

\section*{Appendix B}
\label{3B}
Here we will prove the decomposition of Eq. (\ref{cdcomp}) and relations (\ref{rel6})-(\ref{rel8}). To begin we
will
prove relation (\ref{cdcomp}). Let $\rho({\bf x},t)$ be a
solution of the classical Liouville equation. The following
decomposition can be verified by inspection:
\begin{eqnarray}
\rho({\bf x},t)&=&\int_0^{\infty} dE \int
d{\bf x}_0 ~\rho({\bf x}_0,0)
\frac{\delta(E-H({\bf x}_0))\delta(E-H({\bf x}))}{\int d{\bf
x}'\delta(E-H({\bf x}'))}\nonumber\\
&+&\int_0^{\infty} dE~\rho_{ns}({\bf x},t)\delta(E-H({\bf x})),
\end{eqnarray}
where
\begin{equation}
\rho_{ns}({\bf x},t)=
\rho({\bf x},t) -\langle\rho\rangle_E,
\label{gmess}
\end{equation}
with $\langle\rho\rangle_E$ as given in Eq. (\ref{mcav}).
Furthermore, we may write
\begin{eqnarray}
\rho_{ns}({\bf x},t)&=&\int
d{\bf x}_0 ~\rho_{ns}({\bf x}_0,0)\delta({\bf x}_0-{\bf
X}({\bf x},-t))
\nonumber \\
&=&\int_{-\infty}^{\infty} dt'\delta(t'-t)\int
d{\bf x}_0 ~\rho_{ns}({\bf x}_0,0)\delta({\bf x}_0-{\bf
X}({\bf x},-t')).
\label{mess}
\end{eqnarray}
Substituting $\frac{1}{2\pi}\int_{-\infty}^{\infty} d\lambda ~e^{i\lambda(t'-t)}$ for
$\delta(t'-t)$ in Eq. (\ref{mess}), inserting $1=\int_0^{\infty}dE~\delta(E-H({\bf x}))$, and then substituting
Eq. (\ref{mess}) into Eq. (\ref{gmess}) and collecting terms, we obtain
\begin{eqnarray}
\rho({\bf x},t)&=&\int_0^{\infty} dE\int
d{\bf x}_0 ~\rho({\bf x}_0,0)\Upsilon_E({\bf x};{\bf x}_0)\nonumber\\
&+&\int_0^{\infty} dE\int\!\!\!\!\!\!- d\lambda e^{-i\lambda t}\int
d{\bf x}_0~\rho({\bf x}_0,0)~\Upsilon_{E,\lambda}({\bf x};{\bf x}_0),
\label{dc}
\end{eqnarray}
where $\int\!\!\!\!\!- d\lambda~\rho=\int d\lambda ~P_E~\rho=\int
d\lambda~[\rho-\langle\rho\rangle_E]$, which proves Eq. (\ref{cdcomp}).
Since Eq. (\ref{dc}) must hold for all $t$ it must also hold for $t=0$.
We thus have shown that
\begin{equation}
\rho({\bf x},0)=\int_0^{\infty} dE\int
d{\bf x}_0 ~\rho({\bf x}_0,0)\Upsilon_E({\bf x};{\bf x}_0)+\int_0^{\infty} dE\int\!\!\!\!\!\!- d\lambda\int
d{\bf x}_0~\rho({\bf x}_0,0)~\Upsilon_{E,\lambda}({\bf x};{\bf x}_0),
\end{equation}
and since this must hold for all $\rho$, relation (\ref{rel6}) must be
true. Here of course we have used definitions (\ref{d1}) and (\ref{d2}).

The proof of relation (\ref{rel7}) is much simpler.
Inserting definition (\ref{d1}) into the left-hand side of relation
(\ref{rel7}) gives
\begin{eqnarray}
&&\int d{\bf x}~\Upsilon_{E^{\prime}}^{*}({\bf x};{\bf
x}_0^{\prime})\Upsilon_{E}({\bf x};{\bf x}_0)=
\int d{\bf x}~\frac{\delta(E'-H({\bf x}_0'))\delta(E'-H({\bf x}))}{[\int d{\bf
x}'\delta(E-H({\bf x}'))]^2}\cdot\nonumber \\
&&\delta(E-H({\bf x}_0))\delta(E-H({\bf x}))\nonumber \\
&&=\delta(E-E')\frac{\delta(E-H({\bf x}_0'))\delta(E-H({\bf
x}_0))}{\int d{\bf
x}'\delta(E-H({\bf x}'))}.
\end{eqnarray}
Simple manipulations and use of definition (\ref{d1}) then give Eq.
(\ref{rel7}). 

Relation (\ref{rel8}) can be proven as follows. From definition
(\ref{d2}) it follows that
\begin{eqnarray}
&&\int d{\bf x}~\Upsilon_{E^{\prime},\lambda^{\prime}}^*({\bf x};{\bf x}_0^{\prime})\Upsilon_{E,\lambda}({\bf x};{\bf x}_0)=\nonumber \\
&&\frac{1}{(2\pi)^2}\delta(E-E^{\prime})\int
d{\bf x}\int_{-\infty}^{\infty}dtdt^{\prime}~\delta(E-H({\bf x}))e^{-i\lambda
t}e^{i\lambda^{\prime}t^{\prime}}\cdot\nonumber \\
&&\delta({\bf x}_0^{\prime}-{\bf X}({\bf x},-t))
\delta({\bf x}_0-{\bf X}({\bf x},-t^{\prime}))
\end{eqnarray}
and this can be rewritten in the form
\begin{eqnarray}
&&\int d{\bf x}~\Upsilon_{E^{\prime},\lambda^{\prime}}^*({\bf x};{\bf x}_0^{\prime})\Upsilon_{E,\lambda}({\bf x};{\bf x}_0)=\nonumber \\
&&\frac{1}{(2\pi)^2}\delta(E-E^{\prime})\int_{-\infty}^{\infty}dtdt^{\prime}\delta(E-H({\bf x}_0))e^{-i\lambda
t}e^{i\lambda^{\prime}t^{\prime}}\cdot\nonumber \\
&&\delta({\bf x}_0-{\bf X}({\bf x}_0^{\prime},-(t^{\prime}-t)))
\end{eqnarray}
and now changing $t^{\prime}\rightarrow t+t^{\prime}$ we readily
obtain Eq. (\ref{rel8}).

\pagebreak


\begin{thebibliography}{99}

\bibitem{chacor} J. Wilkie and P. Brumer, following paper.  

\bibitem{cp} L.D. Landau and E.M. Lifshitz,
{\em Quantum Mechanics, Nonrelativistic Theory}, (Pergamon Press,
London, 1958); E. Merzbacher,
{\em Quantum Mechanics}, (Wiley, New York, 1961).

\bibitem{ford2} J. Ford in {\em Directions in Chaos, Vol. 2}, ed. Hao
Bai-Lin (World Scientific, Singapore, 1988).

\bibitem{ford} J. Ford, G. Mantica and G.H. Ristow, Physica D 50, 493
(1991); J. Ford and M. Ilg, Phys. Rev. A 45, 6165 (1992).

\bibitem{berry4} M.V. Berry, in {\em Chaos and Quantum Physics}, ed. M.J.
Giannoni, A. Voros, J. Zinn-Justin, (North-Holland, Amsterdam, 1991).

\bibitem{bw1} J. Wilkie and P. Brumer, Phys. Rev. E 49, 1968 (1994).


\bibitem{Koopman} B.O. Koopman, Proc. Natl. Acad. Sci. U.S.A.,
17, 315 (1931).

\bibitem{dirac} P.A.M. Dirac,
{\em The Principles of Quantum Mechanics}, (Oxford University Press,
Oxford, 1958).

\bibitem{dirac2} This approach is entirely consistent with Dirac's
approach to the reverse problem, i.e., quantization, where the Poisson
bracket is replaced by the commutator.

\bibitem{jaffe2}
C. Jaff\'{e} and P. Brumer, J. Chem. Phys. 82, 2330 (1985).

\bibitem{jaffe1}
C. Jaff\'{e} and P. Brumer, J. Phys. Chem. 88, 4829 (1984).

\bibitem{arnold} V.I. Arnold and A. Avez,
{\em Ergodic Problems of Classical Mechanics}, (Addison-Wesley, N.Y.
1989).

\bibitem{ww} S.R. De Groot and L.G. Suttorp,
{\em Foundations of Electrodynamics}, (North-Holland, Amsterdam,
1972).

\bibitem{necnsuf} This requirement is necessary but not sufficient to 
completely characterize the classical eigenvalue problem, since spectral
degeneracies {\em not} associated with a classical constant of the motion
occur for chaotic systems\cite{arnold}. See Sec. \ref{CS1}.

\bibitem{berry1} M.V. Berry, Phil. Trans. R. Soc. Lond. A 287, 237
(1977).

\bibitem{delos} I.C. Percival, Adv. Chem. Phys. 36, 1 (1977); J.B. Delos, Adv. Chem. Phys. 65, 161 (1986).

\bibitem{carvalho} For an example of the confusion surrounding
classical Liouville eigenfunctions for chaotic systems see our comment
[J. Wilkie and P. Brumer, Phys. Rev. Lett., (submitted)]
on the work of T.O. de Carvalho and M.A.M. de Aguiar, Phys. Rev. Lett. 76,
2690 (1996).

\bibitem{berry2} M.V. Berry, J. Phys. A 10, 2083 (1977).

\bibitem{voros} A. Voros, in {\em Stochastic Behavior in Classical and
Quantum Hamiltonian Systems}, ed. G. Casati and J. Ford, (Springer,
Berlin, 1979).

\bibitem{berry3} M.V. Berry, Proc. R. Soc. Lond. A 423, 219,
(1989).

\bibitem{spec} Our use of the terms point and continuous spectrum and their tacit
association with $L^2$ and singular functions respectively is in
accord with Dirac's use of improper functions and consistent with
current usage in scattering theory.  

\bibitem{footnote2} To see this,
consider that $\omega_1,\omega_2,\dots,\omega_s$ commensurate implies
that $\mbox{\boldmath $\omega$}=\frac{2\pi}{T({\bf I}')}{\bf k}({\bf
I}')$ where ${\bf k}({\bf I}')\in {\bf Z}^s$. Now note that any
orbit originating from $({\bf I}', \mbox{\boldmath $\theta$}')$ at
$t=0$ passes through the point
\begin{eqnarray}
({\bf I}', \mbox{\boldmath $\theta$}'(T({\bf I}'))&=&({\bf I}',
(\mbox{\boldmath $\omega$}T({\bf I}')+\mbox{\boldmath
$\theta$}')~{\rm mod}~2\pi)\nonumber \\
&=&({\bf I}',
(2\pi {\bf k}({\bf I}')+\mbox{\boldmath
$\theta$}')~{\rm mod}~2\pi)\nonumber \\
&=&({\bf I}', \mbox{\boldmath $\theta$}')
\end{eqnarray}
at time $t=T({\bf I}')$. Thus every orbit is periodic.

\bibitem{footnote3} Tori for which
the frequencies $\omega_1,\omega_2,\dots,\omega_s$ are incommensurate
are constructed from non-periodic orbits. This can be seen by
supposing that a point $({\bf I}', \mbox{\boldmath $\theta$}')$ of the
torus exists such that for some time $t=T$, 
\begin{equation}
\mbox{\boldmath $\theta$}'(T)=(\mbox{\boldmath $\omega$}T+\mbox{\boldmath
$\theta$}')~{\rm mod}~2\pi=\mbox{\boldmath $\theta$}'.
\label{contra}
\end{equation}
Equation (\ref{contra}) implies that there exists an ${\bf l}\in {\bf
Z}^s$ such that $T=\frac{2\pi l_j}{\omega_j}$ for all nonzero
components $\omega_j$ of $\mbox{\boldmath $\omega$}$. But this implies
that $\omega_1,\omega_2,\dots,\omega_s$ are commensurate, in
contradiction with our assumption. Hence all orbits of an
incommensurate torus are non-periodic.

\bibitem{footnote4}  Jaffe and Brumer (Ref. \cite{jaffe1}) had previously
proposed these as Liouville eigenfunctions for chaotic systems.

\bibitem{gold} H. Goldstein, {\em Classical Mechanics},
(Addison-Wesley, New York, 1950).

\bibitem{vlad} V.S. Vladimirov, {\em Generalized Functions in Mathematical
Physics}, (Mir, Moscow, 1979).

\bibitem{hannay} J.H. Hannay and A.M. Ozorio de Almeida, J. Phys. A
17, 3429 (1984).

\end{thebibliography}
\end{document}